\documentclass[%
twocolumn,showpacs,preprintnumbers,prb,aps,psfig]{revtex4-2}
\usepackage{graphicx}
\usepackage{dcolumn}
\usepackage{bm}

\usepackage[utf8]{inputenc}
\usepackage{mathptmx}
\usepackage{etoolbox}

\makeatother

\usepackage[dvipsnames]{xcolor}
\usepackage{hyperref}
\hypersetup{
	colorlinks,
	linkcolor=blue,
	citecolor={blue},
	urlcolor={blue}
}

\usepackage{amsmath}
\begin{document}

\title{Orbital selective Mott transition and magnetic moment in charge density wave heterostructures NbSe$_2/$Ta$X_2$}

\date{\today}

\thanks{$^\P$These authors contributed equally.}
\author{Joydeep Chatterjee$^\P$}\email{joydeepch2010@gmail.com}
\author{Shubham Patel$^\P$}\email{spatelphy@iitkgp.ac.in}
\author{A Taraphder}\email{arghya@phy.iitkgp.ac.in}

\affiliation{Department of Physics, Indian Institute of Technology, Kharagpur-721302, India}

\begin{abstract}
 We investigate the electronic properties of charge density wave (CDW) heterostructures out of monolayers of 1T-NbSe$_2$ and 1T-Ta$X_2$ (where, $X=$ S and Se) using first-principles followed by dynamical calculations. The CDW-ordered crystal structures are simulated using $\sqrt{13}\times\sqrt{13}$ supercells of NbSe$_2$ and Ta$X_2$. These two-dimensional heterostructures are modeled by stacking monolayers of NbSe$_2$ and Ta$X_2$ along (001) direction. Our investigations reveal the presence of non-zero magnetic moments in NbSe$_2/$TaS$_2$, albeit without a long-range magnetic order, raising the issue of a possible quantum spin liquid (QSL) as suggested for monolayer 1T-TaS$_2$ recently. In contrast, the NbSe$_2/$TaSe$_2$ heterostructure exhibits no magnetic moment. In order to capture the dynamical effects of local correlation, we use DFT plus multi-orbital dynamical mean field theory (MO-DMFT). Our findings indicate that NbSe$_2/$TaS$_2$ is considerably influenced by the dynamic corrections, whereas NbSe$_2/$TaSe$_2$ shows minimal effects. Additionally, an orbital-selective Mott transition (OSMT) is observed in the NbSe$_2/$TaS$_2$ bilayer heterostructure.
\end{abstract}

\maketitle

\section{\label{sec:intro}  Introduction}

Flat bands in electronic structures indicate a significant localization of electrons, where the on-site Coulomb interaction ($U$) dominates over the bandwidth ($t$), i.e., $U/t>1$, leading to a Mott transition due to electron correlations. Materials such as NbSe$_2$, TaS$_2$, and TaSe$_2$ exhibit metallic, dispersive bands in their pristine bulk and monolayer forms. Notably, the restructured $\sqrt{13}\times\sqrt{13}$ star-of-David (SoD) clusters in 1T-TaS$_2$ and 1T-TaSe$_2$ host these flat bands \cite{fazekas1979electrical, fazekas1980charge, smith1985band, rossnagel2006spin}. Monolayer SoD structures in 1T-NbSe$_2$ are also studied for their flat band and Mott insulation properties. ARPES measurements confirm the presence of a robust flat band persisting from 40K up to 450K \cite{nakata2021robust}. For 1T-TaSe$_2$ and 1T-NbSe$_2$, ARPES studies reveal Mott \textit{half}-gaps of 0.28 eV and 0.23 eV, respectively \cite{nakata2021robust}. 1T-TaS$_2$, recognized as a Mott insulator, exhibits Mott gaps below the Fermi level ranging from 0.2 to 0.4 eV in photoelectron spectroscopy and ARPES experiments \cite{dardel1992temperature, perfetti2003spectroscopic, perfetti2006time, lahoud2014emergence, qiao2017mottness}. Theoretical investigations using density functional theory (DFT) and dynamical mean field theory (DMFT) calculations consistently support these experimental findings \cite{perfetti2003spectroscopic, perfetti2006time, qiao2017mottness, yu2017electronic, darancet2014three, calandra2018phonon, kamil2018electronic, du2024theoretical}.

In its bulk form, NbSe$_2$ crystallizes in a hexagonal structure and undergoes a charge density wave (CDW) transition at low temperatures \cite{Guster,Malliakas,Calandra,Arguello,Straub}. This CDW phase transition results in a periodic modulation of the charge density, leading to an incommensurate CDW state at higher temperatures and a commensurate CDW state at lower temperatures. When reduced to a monolayer, NbSe$_2$ exhibits intriguing changes in its electronic characteristics due to quantum confinement. Superconductivity in NbSe$_2$ samples is widely investigated \cite{wang2017high,hamill2021two,patel2024electron} and owing to a large spin-orbit coupling (SOC) Ising superconductivity is being looked into, recently \cite{xi2016ising,wickramaratne2020ising,das2023electron}. The competition between CDW and superconductivity is well known in the literature \cite{weber2011extended,ugeda2016characterization,koley2020charge,feng2023dynamical}.

Likewise, TaSe$_2$ and TaS$_2$ have hexagonal lattice structures in their bulk from and are known for their CDW phase transitions at low temperatures and high pressures \cite{Burk,Tsen,freitas2016strong,wang2020band}. The Fermi surface nesting has long been rationalized a consequence for the CDW instability in transition metal dichalcogenides (TMDs) \cite{wilson1975charge,rice1975new,johannes2006fermi}. However, the mechanism has been disregarded in experiments \cite{liu1998momentum,liu2000fermi,dardel1993spectroscopic,straub1999charge,ruzicka2001charge}. A correlated, preformed exciton liquid route is then proposed by Taraphder \textit{et al.}, using local density approximation plus MO-DMFT, in which the CDW state arises as Bose condensation of preformed excitons \cite{taraphder2011preformed,koley2014preformed}.

In addition to the Mott insulator, 1T-TaS$_2$ in its CDW phase also exhibits quantum spin liquid (QSL) \cite{zhou2017quantum} behavior caused by the lack of long range magnetic ordering \cite{law20171t}. The spin liquid state was initially proposed by Anderson and Fazekas for triangular lattice due to magnetic frustration \cite{anderson1973resonating,fazekas1974ground}. The half-filled, localized band centred on the central Ta site of the star of David in the CDW state is likely to have a moment. Recently, the monolayer CDW phase of 1T-NbSe$_2$, has been predicted to be a QSL state, when placed on a $3\times 3$ modulated 1H-NbSe$_2$ monolayer \cite{zhang2024quantum}. The scanning tunneling spectra (STS) of the 1T/1H-NbSe$_2$ heterostructure (HS) shows a sharp Kondo peak at the Fermi level indicating a considerable interaction between spin and charge states in two layers. A similar Kondo peak in STS was also observed earlier for 1T/1H-TaSe$_2$ HS \cite{ruan2021evidence}.

\begin{figure*}[!htb]
	\centering
	\includegraphics[scale=0.65]{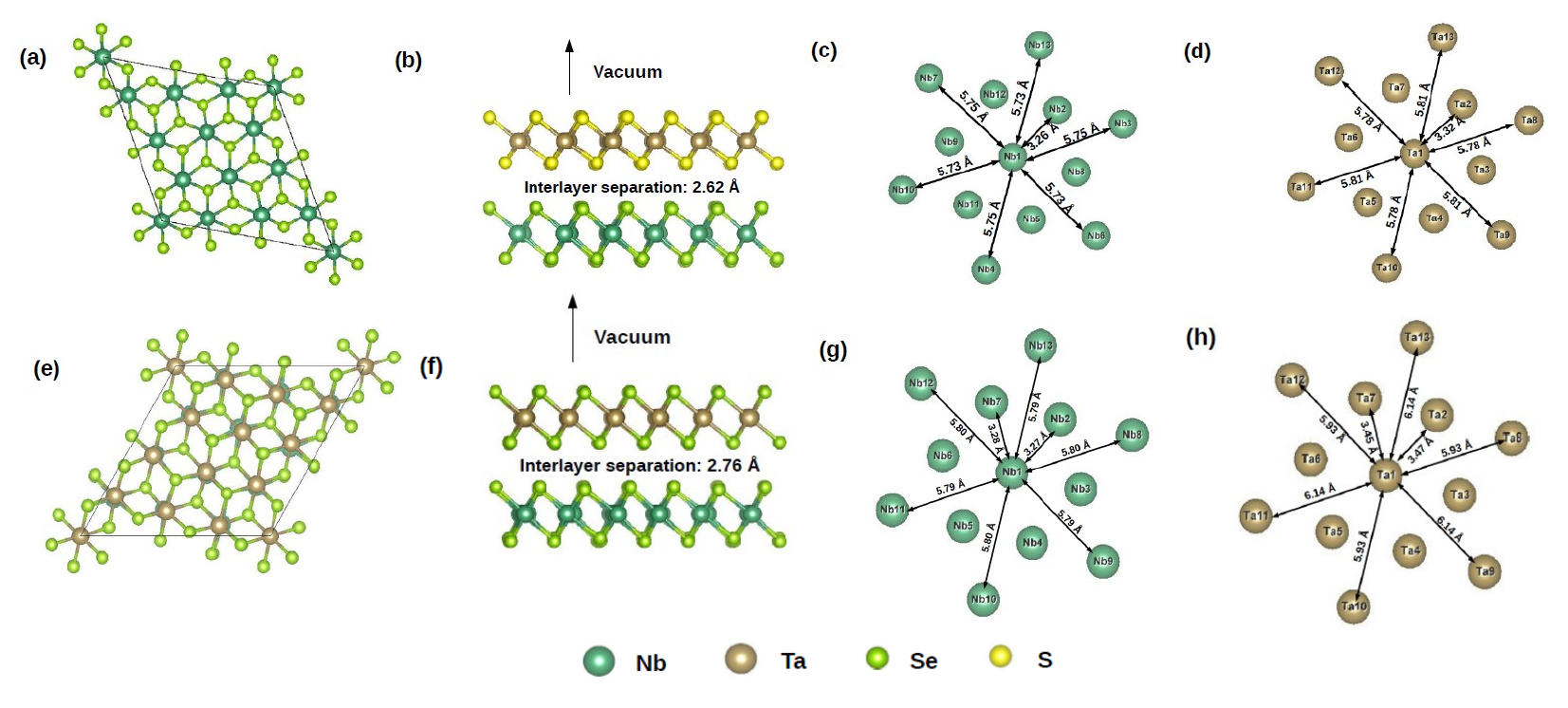}
	\caption{ Crystal structures, (a-d) H1: 1T-NbSe$_2$/TaS$_2$ and (e-h) H2: 1T-NbSe$_2$/TaSe$_2$. (a,e) Top and (b,f) side views. (c,d,g,h) The formation of SoD are shown. The bond lengths are mentioned therein.}
	\label{fig:structs}  
\end{figure*}

The discussion above suggests that the triangular lattices of TMDs are good candidates for the QSL state. By combining different 2D materials with monolayer NbSe$_2$, TaS$_2$ or TaSe$_2$ in HSs, one can create and modulate the quantum states. In the current investigation, we model 2D CDW HSs using 1T polymorph of NbSe$_2$ and Ta$X_2$, where $X=$ Se and S. Our first-principles DFT calculations suggest a possible QSL state in 1T-NbSe$_2$/TaS$_2$ due to the absence of long range magnetic ordering. The magnetic moments on all the Ta/Nb transition metals (TMs) are not uniquely distributed. On the contrary, the QSL state is not observed for 1T-NbSe$_2$/TaSe$_2$ experimentally. We also find no magnetic moment for this heterostructure. Moreover, an orbital selective Mott transition is noted for both the HSs, by incorporating dynamical correlations within DFT+DMFT approach.

\section{\label{sec:comp} Computational Details and Methodology}
\subsection{\label{sec:dft} DFT}
The electronic structures  have been calculated using a plane wave implementation of DFT within the Vienna ab-initio simulation package {\footnotesize (VASP)} \cite{Kresse1,Kresse2,Kresse3,Kresse4} which uses projected augmented wave (PAW) \cite{Blochl,Kresse5} potentials. The generalized gradient approximation \cite{Perdew,Perdew2} is used for the exchange-correlation functional, and dispersion corrections are included within the DFT-D2 approach \cite{Grimme}. Both the lattice parameters and the internal positions are optimized until an energy convergence of $10^{-4}$ eV and force convergence of $10^{-3}$ eV/ {\AA} have been achieved. $\Gamma$-centered k-mesh of 7$\times$7$\times$3 and 8$\times$8$\times$1 has been used for bulk and 2D CDW phases, respectively. A cutoff energy of 520 eV has been used for the plane-wave basis.

\subsection{\label{sec:dmft} DMFT}
We perform the DMFT calculations in a basis set of projective Wannier functions, calculated using {\footnotesize DFTTOOLS} and {\footnotesize soliDMFT} packages \cite{aichhorn2009dynamical,aichhorn2011importance,aichhorn2016triqs,merkel2022soliddmft} interfaced with {\footnotesize TRIQS} libraries \cite{parcollet2015triqs}. The Anderson impurity problems are solved using continuous-time quantum Monte Carlo algorithm in the hybridization ({\footnotesize CTHYB}) expansion \cite{werner2006hybridization} as implemented in {\footnotesize TRIQS} \cite{seth2016triqs}. We use a fully rotationally invariant Kanamori Hamiltonian parametrized by Hubbard's $U$ and Hund's coupling $J_H$, where we set the intraorbital interaction to $U' = U-2J_H$ \cite{vaugier2012hubbard}. For our calculations we used various $U$ values ranging from 1.0 to 3.0 eV and $J_H = 0.8$ eV in order to see the effect of correlations. Real-frequency spectra are obtained using the maximum-entropy method of analytic continuation as implemented in the {\footnotesize TRIQS/MAXENT} application \cite{kraberger2017maximum}. We carry out all the DMFT calculations at an inverse temperature $\beta = 40\, eV^{-1}$.

\begin{figure*}[!htb]
	\centering
	\includegraphics[scale=0.62]{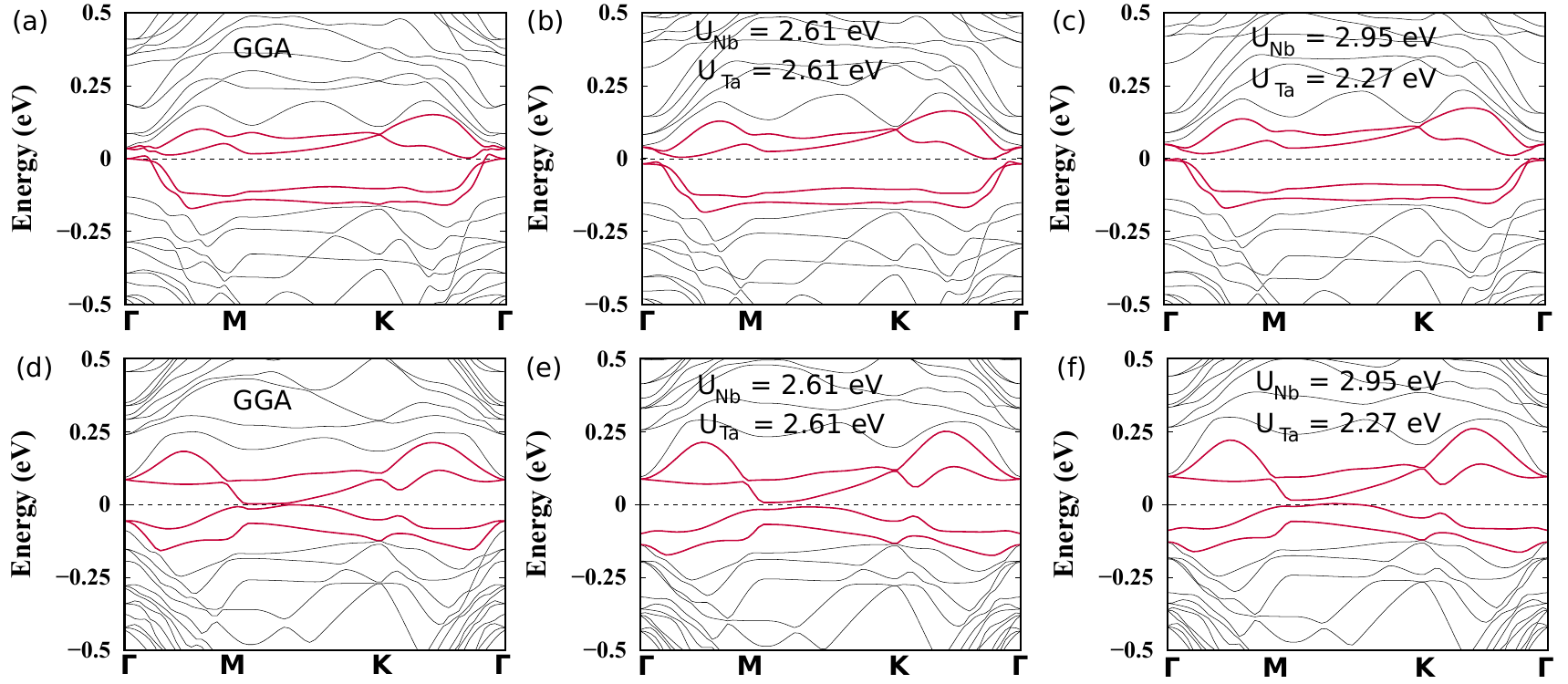}
	\caption{Band structures without spin-polarization of (a-c) 1T-NbSe$_2$/TaS$_2$ and (d-f) 1T-NbSe$_2$/TaSe$_2$ with (a,d) GGA, (b,e) GGA$+U$, where an average value of $U = 2.61$ eV is used, and (c,f) GGA$+U$, where $U_{\rm{Nb}} = 2.95$ eV and $U_{\rm{Ta}} = 2.27$ eV. The red bands are used for Wannier projections. }
	\label{fig:bands_h1h2}  
\end{figure*}

\section{\label{sec:struct} Crystal Structures}

The crystal structures are generated using 1T polymorphs of TMDs, $MX_2$, in which the transition metals (TMs), $M$ are sandwiched between the chalcogens, $X$, in octahedral symmetry. The $\sqrt{13}\times\sqrt{13}$ supercells are created using 13 TM atoms cluster in which 12 TM atoms surround a central TM atom. The surrounding TM atoms are deviated slightly from their original Wyckoff positions and we notice that after relaxation they form SoD like structure. These are well known CDW crystal phases. Next, the HSs are formed by deposition of single layer of SoD structure above another with an interlayer distance, $d_{M-M} \sim 6$\AA. We choose 1T-NbSe$_2$/TaS$_2$ and 1T-NbSe$_2$/TaSe$_2$ HSs for our investigation which we refer H1 and H2, respectively, throughout the paper. For the 2D bilayer HSs a vacuum of 15\AA\, has been added along third direction in order to discard spurious interactions between the periodic unit cells. The crystal structures, interlayer separations between the nearest chlcogens of two layers and the bond lengths are shown in Fig. \ref{fig:structs}.

\begin{center}
	\begin{table}[!htb]
		\centering
		\setlength{\tabcolsep}{16pt}
		\renewcommand{\arraystretch}{1.2}
		\caption{\label{table:ene} Energy per formula unit cell (in eV) for H1 and H2. } 
		\begin{tabular}{ c c c}
			\hline
			\hline
			Structural phase  & for H1 (eV) & for H2 (eV)\\
			\hline
			1T Normal & -22.30 & -21.53\\
			1T CDW & -22.32 & -21.48\\
			
			\hline
			\hline
		\end{tabular}
	\end{table}
\end{center}

In order to check the structural stability of the CDW phase in both the HSs, we calculate energy per formula unit cell ($E_{fu}$) in the normal phase and in the CDW phase. The energies values are provided in Table \ref{table:ene}. $E_{fu}$ is evaluated by calculating the DFT ground state energies of the bulk (containing two $MX_2$ units) and the bilayer CDW structures (containing 26 $MX_2$ units), then dividing the totat energy by the number of $MX_2$ units in that particular system. We find that after the formation of SoD clusters, $E_{fu}$ for H1 becomes slightly more negative than $E_{fu}$ of the normal 1T bulk phase. Conversely, $E_{fu}$ becomes less negative in the case of H2. Therefore, it turns out that the H1 is more stable in its CDW phase while H2 destabilizes in the cluster form. The destabilization of H2 can be attributed to the large bond length differences from the outer Ta atoms to the central Ta atom of SoD as shown in Fig. \ref{fig:structs}(h), which deforms the SoD. The difference in the bond lengths is smaller for H1, unlike H2 (see Fig. \ref{fig:structs}(d)). Additionally, we calculated $E_{fu}$ for the monolayer CDW phases of 1T-NbSe$_2$, 1T-TaS$_2$ and 1T-TaSe$_2$, and the measured values are $-19.59$ eV, $-21.18$ eV, and $-21.25$ eV, respectively. Though the direct comparision of the energies has no physical meaning the average energy of the two layers is still higher than the bilayer system, indicating the bilayers HSs are more stable than the monolayers. The Van der Waals interactions could be playing a role in stabilizing the bilayer systems.

\section{\label{sec:results}  Results and Discussions}	

\subsection{\label{sec:dftresults}  DFT, DFT$\mathbf{+\textit{U}}$ and spin-polarization}
We calculate the band structures for H1 (Fig. \ref{fig:bands_h1h2}(a-c)) and H2 (Fig. \ref{fig:bands_h1h2}(d-f)) under GGA and GGA+$U$ formalisms. We find that without on-site Hubbard $U$, the highest valence band (HVB) crosses the Fermi level (FL) along $\Gamma-L$, and the lowest conduction band (LCB) merely touch the FL in case of H1 (see Fig. \ref{fig:bands_h1h2}(a)). In case of H2, the HVB and LCB do not cross but touch the FL as shown in Fig. \ref{fig:bands_h1h2}(d). However, the HVB and LCB cross or touch the FL along differnt momentum directions in H1 and H2. For H1, the HVB cross the FL along $\Gamma-M$ and $K-\Gamma$, whereas the momentum direction is $M-K$ for H2, where the bands touch the FL. The HVB in H1 forms tiny hole pockets along $\Gamma-M$ and $K-\Gamma$ Brillouin zone (BZ) directions. In rest of the BZ, the bands are nearly flat, indicating strong correlations supported by the large on-site interactions and small kinetic energy between the electrons. The band widths of HVBs for H1 and H2 are $0.12$ and $0.09$ eV, respectively. Our atom and orbital-projected density of states (DOS) calculations suggests that the four bands near the FL (highlighted in red in Fig. \ref{fig:bands_h1h2}) corresponds to the central Nb (Ta) $4d$ ($5d$) orbitals. Moreover, there is a strong hybridization of different $d$ orbitals.

In order to see the effect of static on-site Hubbard $U$ on electron localization, we perform GGA$+U$ calculations, with different $U$ values for those having $d$ orbitals, Ta and Nb TM atoms in our case. As shown in the band structure for H1, $U_{\rm{Nb/Ta}} = 2.61$ eV has no significant effect on the overall band topology, except, the two valence bands close to the FL are shifted down by a negligible amount of $0.012$ eV. On the other hand if we use different values of $U$ for Ta ($U=2.95$ eV) and Nb ($U=2.27$ eV), the two conduction bands slightly move up in the energy with respect to the GGA bands, and the two valence bands shift down, only touching the FL, without crossing it. The hole pockets in this case disappear. The band gap values achieved for H1 with $U_{\rm{Nb/Ta}} = 2.61$ and $U_{\rm{Nb\,(Ta)}} = 2.95\,(2.27)$ eV are $9.84$ meV and $10.5$ meV, respectively. It is important to emphasize here that the monolayer 1T CDW phase of NbSe$_2$ \cite{calandra2018phonon,liu2021direct} and TaS/Se$_2$ \cite{nakata2021robust} are considered to be Mott insulators, with $U=2.8$ eV. The Mott insulation phase is also achieved in the case of HSs but rather small band gap values. Though we have not checked, following the discussion above regarding varying $U$ and shifting bands, we believe a band gap can be induced if we increase $U$ for Ta. On the other hand, a band gap of $14.89$ meV is opened with $U_{\rm{Nb/Ta}} = 2.61$ eV in the case of H2 (see Fig. \ref{fig:bands_h1h2}(e)). Similar to H1, the LCB shifts up with $U_{\rm{Nb}} = 2.95$ eV and $U_{\rm{Ta}} = 2.27$ eV (see Fig. \ref{fig:bands_h1h2}(f)).

\begin{figure}[!htb]
	\centering
	\includegraphics[scale=0.30]{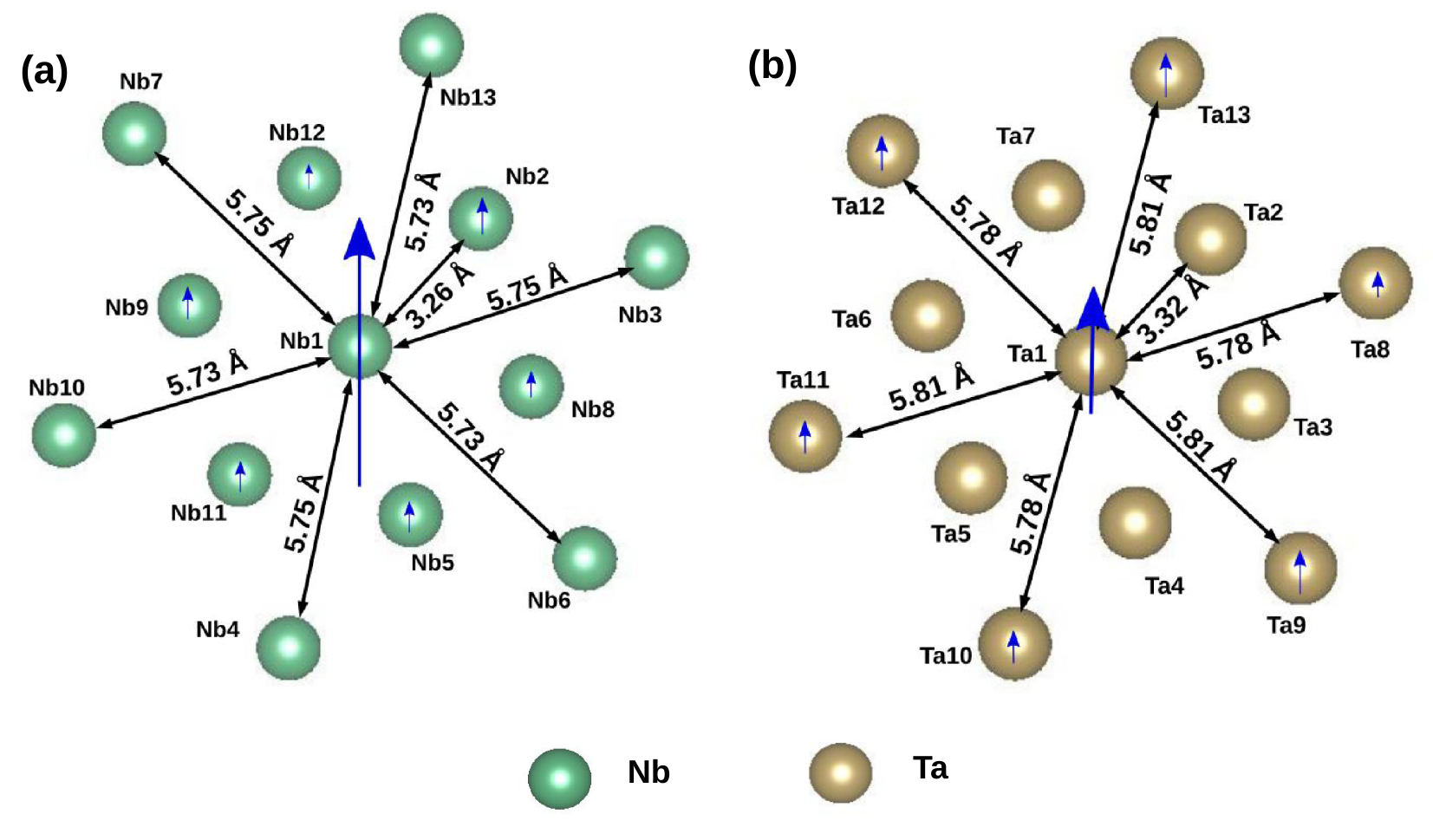}
	\caption{The magnitudes of magnetic moments in respective layers of stars of H1 are shown by the lengths of arrows with $U_{\rm{Nb}} = 2.95$ eV and $U_{\rm{Ta}} = 2.27$ eV. }
	\label{fig:stars}  
\end{figure}

As discussed in the introduction, monolayer 1T-TaS$_2$ exhibit a QSL state, which regards a lack of long range magnetic ordering. In the following, we discuss the magnetic structure of the bilayer HSs. We perform spin-polarized DFT and DFT$+U$ calculations, and find that the 1T CDW phase comes out to be non-magnetic without $U$ for both the HSs. Incorporating $U$ in the calculations induce magnetic moment on TM atoms for the H1 HS, and most of the contribution of magnetic moment is concentrated on the central Ta and Nb atoms, as denoted by arrows in Fig. \ref{fig:stars}. The size of the arrow depicts the magnitude of the magnetic moment, which is maximum for the central atoms of SoD. The magnetic moment of the central Nb and Ta, with $U_{\rm{Nb}} = 2.95$ eV and $U_{\rm{Ta}} = 2.27$ eV, are $0.684$ and $0.311$ $\mu_B$, respectively, where, $\mu_B$ is Bohr magneton. It is interesting to see that in the SoD of NbSe$_2$ layer, the nearest Nb atoms from the central Nb show the secondary contribution in the magnetic moment while for the TaS$_2$ layer, the secondary contribution is dominated from the next nearest neighbors. The secondary contribution in the magnetic moments is of the order of $10^{-2}$ $\mu_B$, while the lowest contribution is either zero or of the order of $10^{-3}$ $\mu_B$. The magnetic moment on each species is not uniquely defined, causing the absence of long range magnetic ordering in a particular layer as well as in the bilayer, which might result into a QSL state. We perform spin-polarized calculations for $U_{\rm{Nb/Ta}} = 2.61$ eV, and found that only central atoms dominantly contribute to the magnetic moment with the values, $0.117$ $\mu_B$ and $-0.064$ $\mu_B$ for Nb and Ta, respectively. The rest of the TMs exhibit the magnetic moment of the order of $10^{-3}$ $\mu_B$ or zero. Here, the negative value of moment for Ta represents spin down. Furthermore, the spin-polarized band structures clearly show spin-split bands with non-zero $U$ (not shown). Additionally, the spin-polarized bands of H1 with $U_{\rm{Nb/Ta}} = 2.61$ eV show a band gap of $8.34$ meV, while the system is metallic with $U=0$ and $U_{\rm{Nb\,(Ta)}} = 2.95\,(2.27)$ eV. H2 HS, on the contrary, is non-magnetic with and without $U$, and there is no spin-plitting in the spin-polarized bands. As reported earlier, and mentioned in the introduction, 1T-TaSe$_2$ shows the signature of QSL with 1T-TaS$_2$ \cite{chen2020strong,ruan2021evidence}, 1T-TaS$_2$ itself being a QSL \cite{law20171t}. Evidently, our investigation finds that the HS containing 1T-TaS$_2$ (H1) could be a potential candidate for QSL, unlike the H2.

\subsection{\label{sec:dmftresults}  DMFT calculations}
In order to see the electron dynamics we perform DMFT calculations. It is evident from the DFT band structure calculations that H1 and H2 exhibit flat bands near the FL, postulating strong correlations effects in the electrons. Strong correlation can induce Mott insulating behavior in 2D limit. We extract four bands around the FL by adopting the maximally localized Wannier functions (MLWFs) \cite{marzari1997maximally} using Wannier90 package \cite{pizzi2020wannier90} as implemented in {\footnotesize QUANTUM ESPRESSO} (QE) \cite{giannozzi2009quantum,giannozzi2017advanced,giannozzi2020quantum}. As discussed above in Sec. \ref{sec:dftresults}, there are multiple 4$d$ and 5$d$ orbitals hybridizing near the FL. Our extracted Wannier bands are nearly good match with DFT bands, and also the spreads of Wannier functions are small. The slight mismatch can be attributed to the orbital-hybridization, and considering only four most contributing orbitals near the FL. We choose $4d_{xy}$, $4d_{z^2}$, $5d_{xy}$ and $5d_{z^2}$ for Wannier fitting. Different values of Hubbard parameter, $U$, in the fully rotationally invariant Kanamori Hamiltonian \cite{vaugier2012hubbard} are used to see the variations in the bands and to investigate the Mott insulating phase. We would like to mention that there is a little mismatch in the Fermi energy calculated using {\footnotesize VASP} and QE. Owing to that, the valence bands extracted with Wannier90, shifts up in the energy and cross the FL, which we are going to discuss shortly. Also, $U$ is not taken into account in the DFT bands from which the Wannier bands are extracted. 

\begin{figure*}[!htb]
	\centering
	\includegraphics[scale=0.08]{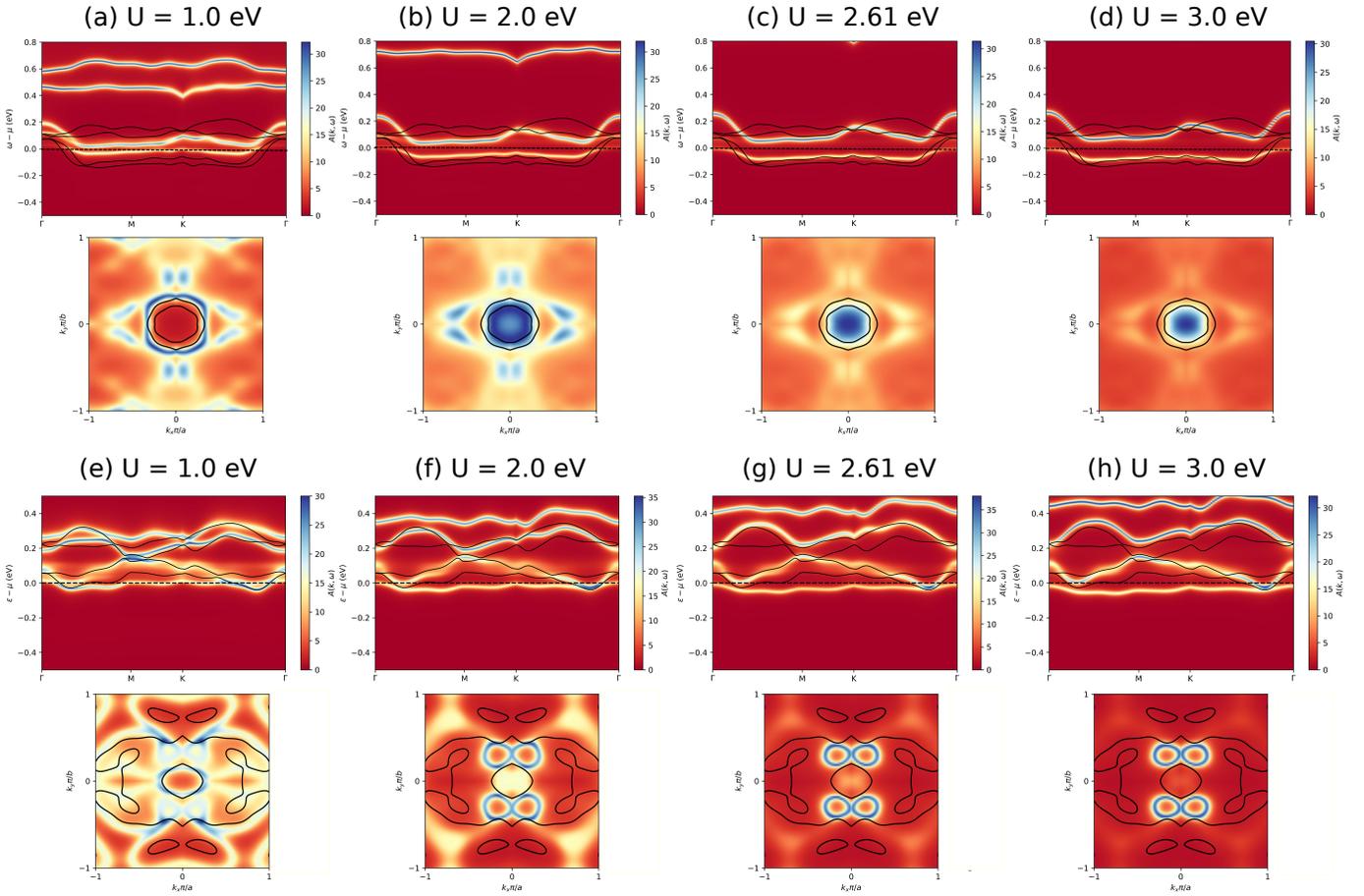}
	\caption{ Upper panel: spectral functions with different values of $U$, and bottom panel: the corresponding Fermi surfaces for (a-d) H1 and (e-h) H2. The dashed line denotes the FL.}
	\label{fig:dmft_bands}  
\end{figure*}

The momentum resolved spectral function (DMFT spectrum) is shown for H1 in Fig. \ref{fig:dmft_bands}(a-d) and for H2 in Fig. \ref{fig:dmft_bands}(e-h). The DFT bands are shown in black while the bright bands are spectral functions. As it can be seen that for H1 the lowest two conduction bands which are called upper Hubbard bands (UHBs) in the literature move significantly up with $U=1$ eV (see Fig. \ref{fig:dmft_bands}(a)) and separate from the highest two valence bands which are called lower Hubbard bands (LHBs). With increasing $U$ the two UHBs goes up in the energy very rapidly from $U=1$ to $U=3$ eV. Notably, the LHBs also start isolating from each other at $U=2$ eV (see Fig. \ref{fig:dmft_bands}(b)) and at $U=2.61$ eV (see Fig. \ref{fig:dmft_bands}(c)) they completely detach and open up a gap in the spectrum. This gap opening is a signature of orbital-selective Mott transition (OSMT). OSMT is defined as a phase in which some of the orbitals show strong correlations (localized electrons) while the others remain itinerant \cite{anisimov2002orbital,vojta2010orbital}. The OSMT is robust with higher $U$ values (see Fig. \ref{fig:dmft_bands}(d)). In addition to OSMT, splitting in the DMFT bands can be seen at the $\Gamma$ point, unlike the DFT bands. A weak satellite feature is also apparent near the $\Gamma$ point with $U=1$ and $2$ eV in the lower VHB. The spectral weight is isolated from the original band around $\Gamma$ point opening a gap. The satellite features in the spectral function are previously reported for cubic SrVO$_3$ \cite{boehnke2016when}. These satellites are signature of weak localization.

The OSMT is also evident from the Fermi surface (FS) shown in the bottom panels of Fig. \ref{fig:dmft_bands}(d). As presented in Figs. \ref{fig:dmft_bands}(a-d), the distribution of spectral function is denoted by blue and yellow intensities. The intensities start disappearing towards higher $U$ values and almost vanish for $U=3$ eV, except, a smaller and intense hole pocket around the $\Gamma$ point. The FSs further support OSMT in H1. On the other hand, H2 neither exhibit OSMT nor the satellites. As apparent from Figs. \ref{fig:dmft_bands}(e-h) the HVB (or LHB) always intersect the FL. Moreover, a shifting in UHBs to the higher energy is visible in H2 also with $U$, but the sensitivity with $U$ is smaller as compared to the H1 HS. The LHBs, unlike UHBs, are almost unaltered with $U$ and show no shifting.

\section{\label{sec:conclusions}  Conclusions}
In conclusion, the CDW bilayer phases of 1T-NbSe$_2$/TaS$_2$ (H1) and 1T-NbSe$_2$/TaSe$_2$ (H2) are investigated for their structural and electronic properties. It is found that after relaxation H1 form a SoD structure, while the SoD is deformed in case of H2, leading to a unstable CDW structure of H2 as compared to its bulk phase. The ground state energy per formula unit ($E_{fu}$) suggests that H1 is stable in its CDW phase with a $0.02$ eV lesser $E_{fu}$. Also, the bilayer HSs are found to be stable than their monolayer counterparts. The band structure calculations presents flat bands in the spectrum indicating strong correlations due to electron localization. We observe metallic (H1) and nearly insulating (H2) states without $U$. With non-zero $U$ band gaps is produced in both the heterostructures resulting into Mott insulators. We also report a possible QSL behavior in H1 which we explain by non-zero and random magnetic moments on TM atoms. A QSL state is previously observed experimentally in monolayer CDW phase of 1T-TaS$_2$ and 1T-TaS$_2$-based bilayers systems. Unlike H1, in H2, we do not find magnetic moments on TMs with or without $U$, indicating QSL is uniquely defined for 1T-TaS$_2$-based HSs. Moreover, our DMFT calculations predict OSMT in H1, and also satellite features in the spectral function. The OSMT and satellites are absent in H2 HS. We find a significant sensitivity to $U$ in the spectral function for H1, unlike H2 that shows a very slow variation by varying $U$. The LHBs in H2 are almost fixed and hardly show any variation at $U \geq 2$ eV.

\section{\label{sec:ackn}Acknowledgements}
AT and JC acknowledge funding from SERB (Government of India) CRG project No: CRG/2021/001078 under which this work was performed. SP thanks Alexander Hampel for technical help and useful discussions regarding TRIQS software. The authors acknowledge National Supercomputing Mission (NSM) for providing computing resources of 'PARAM Shakti' at IIT Kharagpur, which is implemented by C-DAC and supported by the Ministry of Electronics and Information Technology (MeitY) and Department of Science and Technology (DST), Government of India. 

\bibliography{ref}

\begin{thebibliography}{78}%
\makeatletter
\providecommand \@ifxundefined [1]{%
 \@ifx{#1\undefined}
}%
\providecommand \@ifnum [1]{%
 \ifnum #1\expandafter \@firstoftwo
 \else \expandafter \@secondoftwo
 \fi
}%
\providecommand \@ifx [1]{%
 \ifx #1\expandafter \@firstoftwo
 \else \expandafter \@secondoftwo
 \fi
}%
\providecommand \natexlab [1]{#1}%
\providecommand \enquote  [1]{``#1''}%
\providecommand \bibnamefont  [1]{#1}%
\providecommand \bibfnamefont [1]{#1}%
\providecommand \citenamefont [1]{#1}%
\providecommand \href@noop [0]{\@secondoftwo}%
\providecommand \href [0]{\begingroup \@sanitize@url \@href}%
\providecommand \@href[1]{\@@startlink{#1}\@@href}%
\providecommand \@@href[1]{\endgroup#1\@@endlink}%
\providecommand \@sanitize@url [0]{\catcode `\\12\catcode `\$12\catcode
  `\&12\catcode `\#12\catcode `\^12\catcode `\_12\catcode `\%12\relax}%
\providecommand \@@startlink[1]{}%
\providecommand \@@endlink[0]{}%
\providecommand \url  [0]{\begingroup\@sanitize@url \@url }%
\providecommand \@url [1]{\endgroup\@href {#1}{\urlprefix }}%
\providecommand \urlprefix  [0]{URL }%
\providecommand \Eprint [0]{\href }%
\providecommand \doibase [0]{https://doi.org/}%
\providecommand \selectlanguage [0]{\@gobble}%
\providecommand \bibinfo  [0]{\@secondoftwo}%
\providecommand \bibfield  [0]{\@secondoftwo}%
\providecommand \translation [1]{[#1]}%
\providecommand \BibitemOpen [0]{}%
\providecommand \bibitemStop [0]{}%
\providecommand \bibitemNoStop [0]{.\EOS\space}%
\providecommand \EOS [0]{\spacefactor3000\relax}%
\providecommand \BibitemShut  [1]{\csname bibitem#1\endcsname}%
\let\auto@bib@innerbib\@empty
\bibitem [{\citenamefont {Fazekas}\ and\ \citenamefont
  {Tosatti}(1979)}]{fazekas1979electrical}%
  \BibitemOpen
  \bibfield  {author} {\bibinfo {author} {\bibfnamefont {P.}~\bibnamefont
  {Fazekas}}\ and\ \bibinfo {author} {\bibfnamefont {E.}~\bibnamefont
  {Tosatti}},\ }\bibfield  {title} {\bibinfo {title} {{Electrical, structural
  and magnetic properties of pure and doped 1T-TaS$_2$}},\ }\href
  {https://doi.org/10.1080/13642817908245359} {\bibfield  {journal} {\bibinfo
  {journal} {Philosophical Magazine B}\ }\textbf {\bibinfo {volume} {39}},\
  \bibinfo {pages} {229} (\bibinfo {year} {1979})}\BibitemShut {NoStop}%
\bibitem [{\citenamefont {Fazekas}\ and\ \citenamefont
  {Tosatti}(1980)}]{fazekas1980charge}%
  \BibitemOpen
  \bibfield  {author} {\bibinfo {author} {\bibfnamefont {P.}~\bibnamefont
  {Fazekas}}\ and\ \bibinfo {author} {\bibfnamefont {E.}~\bibnamefont
  {Tosatti}},\ }\bibfield  {title} {\bibinfo {title} {{Charge carrier
  localization in pure and doped 1T-TaS$_2$}},\ }\href
  {https://doi.org/https://doi.org/10.1016/0378-4363(80)90229-6} {\bibfield
  {journal} {\bibinfo  {journal} {Physica B+C}\ }\textbf {\bibinfo {volume}
  {99}},\ \bibinfo {pages} {183} (\bibinfo {year} {1980})}\BibitemShut
  {NoStop}%
\bibitem [{\citenamefont {Smith}\ \emph {et~al.}(1985)\citenamefont {Smith},
  \citenamefont {Kevan},\ and\ \citenamefont {DiSalvo}}]{smith1985band}%
  \BibitemOpen
  \bibfield  {author} {\bibinfo {author} {\bibfnamefont {N.}~\bibnamefont
  {Smith}}, \bibinfo {author} {\bibfnamefont {S.}~\bibnamefont {Kevan}},\ and\
  \bibinfo {author} {\bibfnamefont {F.}~\bibnamefont {DiSalvo}},\ }\bibfield
  {title} {\bibinfo {title} {{Band structures of the layer compounds 1T-TaS$_2$
  and 2H-TaSe$_2$ in the presence of commensurate charge-density waves}},\
  }\href {https://doi.org/10.1088/0022-3719/18/16/013} {\bibfield  {journal}
  {\bibinfo  {journal} {Journal of Physics C: Solid State Physics}\ }\textbf
  {\bibinfo {volume} {18}},\ \bibinfo {pages} {3175} (\bibinfo {year}
  {1985})}\BibitemShut {NoStop}%
\bibitem [{\citenamefont {Rossnagel}\ and\ \citenamefont
  {Smith}(2006)}]{rossnagel2006spin}%
  \BibitemOpen
  \bibfield  {author} {\bibinfo {author} {\bibfnamefont {K.}~\bibnamefont
  {Rossnagel}}\ and\ \bibinfo {author} {\bibfnamefont {N.~V.}\ \bibnamefont
  {Smith}},\ }\bibfield  {title} {\bibinfo {title} {{Spin-orbit coupling in the
  band structure of reconstructed
  $1T\text{\ensuremath{-}}{\mathrm{TaS}}_{2}$}},\ }\href
  {https://doi.org/10.1103/PhysRevB.73.073106} {\bibfield  {journal} {\bibinfo
  {journal} {Phys. Rev. B}\ }\textbf {\bibinfo {volume} {73}},\ \bibinfo
  {pages} {073106} (\bibinfo {year} {2006})}\BibitemShut {NoStop}%
\bibitem [{\citenamefont {Nakata}\ \emph {et~al.}(2021)\citenamefont {Nakata},
  \citenamefont {Sugawara}, \citenamefont {Chainani}, \citenamefont {Oka},
  \citenamefont {Bao}, \citenamefont {Zhou}, \citenamefont {Chuang},
  \citenamefont {Cheng}, \citenamefont {Kawakami}, \citenamefont {Saruta} \emph
  {et~al.}}]{nakata2021robust}%
  \BibitemOpen
  \bibfield  {author} {\bibinfo {author} {\bibfnamefont {Y.}~\bibnamefont
  {Nakata}}, \bibinfo {author} {\bibfnamefont {K.}~\bibnamefont {Sugawara}},
  \bibinfo {author} {\bibfnamefont {A.}~\bibnamefont {Chainani}}, \bibinfo
  {author} {\bibfnamefont {H.}~\bibnamefont {Oka}}, \bibinfo {author}
  {\bibfnamefont {C.}~\bibnamefont {Bao}}, \bibinfo {author} {\bibfnamefont
  {S.}~\bibnamefont {Zhou}}, \bibinfo {author} {\bibfnamefont {P.-Y.}\
  \bibnamefont {Chuang}}, \bibinfo {author} {\bibfnamefont {C.-M.}\
  \bibnamefont {Cheng}}, \bibinfo {author} {\bibfnamefont {T.}~\bibnamefont
  {Kawakami}}, \bibinfo {author} {\bibfnamefont {Y.}~\bibnamefont {Saruta}},
  \emph {et~al.},\ }\bibfield  {title} {\bibinfo {title} {{Robust
  charge-density wave strengthened by electron correlations in monolayer
  1T-TaSe$_2$ and 1T-NbSe$_2$}},\ }\href
  {https://doi.org/10.1038/s41467-021-26105-1} {\bibfield  {journal} {\bibinfo
  {journal} {Nature communications}\ }\textbf {\bibinfo {volume} {12}},\
  \bibinfo {pages} {5873} (\bibinfo {year} {2021})}\BibitemShut {NoStop}%
\bibitem [{\citenamefont {Dardel}\ \emph {et~al.}(1992)\citenamefont {Dardel},
  \citenamefont {Grioni}, \citenamefont {Malterre}, \citenamefont {Weibel},
  \citenamefont {Baer},\ and\ \citenamefont {L\'evy}}]{dardel1992temperature}%
  \BibitemOpen
  \bibfield  {author} {\bibinfo {author} {\bibfnamefont {B.}~\bibnamefont
  {Dardel}}, \bibinfo {author} {\bibfnamefont {M.}~\bibnamefont {Grioni}},
  \bibinfo {author} {\bibfnamefont {D.}~\bibnamefont {Malterre}}, \bibinfo
  {author} {\bibfnamefont {P.}~\bibnamefont {Weibel}}, \bibinfo {author}
  {\bibfnamefont {Y.}~\bibnamefont {Baer}},\ and\ \bibinfo {author}
  {\bibfnamefont {F.}~\bibnamefont {L\'evy}},\ }\bibfield  {title} {\bibinfo
  {title} {{Temperature-dependent pseudogap and electron localization in
  1T-${\mathrm{TaS}}_{2}$}},\ }\href {https://doi.org/10.1103/PhysRevB.45.1462}
  {\bibfield  {journal} {\bibinfo  {journal} {Phys. Rev. B}\ }\textbf {\bibinfo
  {volume} {45}},\ \bibinfo {pages} {1462} (\bibinfo {year}
  {1992})}\BibitemShut {NoStop}%
\bibitem [{\citenamefont {Perfetti}\ \emph {et~al.}(2003)\citenamefont
  {Perfetti}, \citenamefont {Georges}, \citenamefont {Florens}, \citenamefont
  {Biermann}, \citenamefont {Mitrovic}, \citenamefont {Berger}, \citenamefont
  {Tomm}, \citenamefont {H\"ochst},\ and\ \citenamefont
  {Grioni}}]{perfetti2003spectroscopic}%
  \BibitemOpen
  \bibfield  {author} {\bibinfo {author} {\bibfnamefont {L.}~\bibnamefont
  {Perfetti}}, \bibinfo {author} {\bibfnamefont {A.}~\bibnamefont {Georges}},
  \bibinfo {author} {\bibfnamefont {S.}~\bibnamefont {Florens}}, \bibinfo
  {author} {\bibfnamefont {S.}~\bibnamefont {Biermann}}, \bibinfo {author}
  {\bibfnamefont {S.}~\bibnamefont {Mitrovic}}, \bibinfo {author}
  {\bibfnamefont {H.}~\bibnamefont {Berger}}, \bibinfo {author} {\bibfnamefont
  {Y.}~\bibnamefont {Tomm}}, \bibinfo {author} {\bibfnamefont {H.}~\bibnamefont
  {H\"ochst}},\ and\ \bibinfo {author} {\bibfnamefont {M.}~\bibnamefont
  {Grioni}},\ }\bibfield  {title} {\bibinfo {title} {{Spectroscopic Signatures
  of a Bandwidth-Controlled Mott Transition at the Surface of 1T-TaS$_2$}},\
  }\href {https://doi.org/10.1103/PhysRevLett.90.166401} {\bibfield  {journal}
  {\bibinfo  {journal} {Phys. Rev. Lett.}\ }\textbf {\bibinfo {volume} {90}},\
  \bibinfo {pages} {166401} (\bibinfo {year} {2003})}\BibitemShut {NoStop}%
\bibitem [{\citenamefont {Perfetti}\ \emph {et~al.}(2006)\citenamefont
  {Perfetti}, \citenamefont {Loukakos}, \citenamefont {Lisowski}, \citenamefont
  {Bovensiepen}, \citenamefont {Berger}, \citenamefont {Biermann},
  \citenamefont {Cornaglia}, \citenamefont {Georges},\ and\ \citenamefont
  {Wolf}}]{perfetti2006time}%
  \BibitemOpen
  \bibfield  {author} {\bibinfo {author} {\bibfnamefont {L.}~\bibnamefont
  {Perfetti}}, \bibinfo {author} {\bibfnamefont {P.~A.}\ \bibnamefont
  {Loukakos}}, \bibinfo {author} {\bibfnamefont {M.}~\bibnamefont {Lisowski}},
  \bibinfo {author} {\bibfnamefont {U.}~\bibnamefont {Bovensiepen}}, \bibinfo
  {author} {\bibfnamefont {H.}~\bibnamefont {Berger}}, \bibinfo {author}
  {\bibfnamefont {S.}~\bibnamefont {Biermann}}, \bibinfo {author}
  {\bibfnamefont {P.~S.}\ \bibnamefont {Cornaglia}}, \bibinfo {author}
  {\bibfnamefont {A.}~\bibnamefont {Georges}},\ and\ \bibinfo {author}
  {\bibfnamefont {M.}~\bibnamefont {Wolf}},\ }\bibfield  {title} {\bibinfo
  {title} {{Time Evolution of the Electronic Structure of 1T-TaS$_2$ through
  the Insulator-Metal Transition}},\ }\href
  {https://doi.org/10.1103/PhysRevLett.97.067402} {\bibfield  {journal}
  {\bibinfo  {journal} {Phys. Rev. Lett.}\ }\textbf {\bibinfo {volume} {97}},\
  \bibinfo {pages} {067402} (\bibinfo {year} {2006})}\BibitemShut {NoStop}%
\bibitem [{\citenamefont {Lahoud}\ \emph {et~al.}(2014)\citenamefont {Lahoud},
  \citenamefont {Meetei}, \citenamefont {Chaska}, \citenamefont {Kanigel},\
  and\ \citenamefont {Trivedi}}]{lahoud2014emergence}%
  \BibitemOpen
  \bibfield  {author} {\bibinfo {author} {\bibfnamefont {E.}~\bibnamefont
  {Lahoud}}, \bibinfo {author} {\bibfnamefont {O.~N.}\ \bibnamefont {Meetei}},
  \bibinfo {author} {\bibfnamefont {K.~B.}\ \bibnamefont {Chaska}}, \bibinfo
  {author} {\bibfnamefont {A.}~\bibnamefont {Kanigel}},\ and\ \bibinfo {author}
  {\bibfnamefont {N.}~\bibnamefont {Trivedi}},\ }\bibfield  {title} {\bibinfo
  {title} {{Emergence of a Novel Pseudogap Metallic State in a Disordered 2D
  Mott Insulator}},\ }\href {https://doi.org/10.1103/PhysRevLett.112.206402}
  {\bibfield  {journal} {\bibinfo  {journal} {Phys. Rev. Lett.}\ }\textbf
  {\bibinfo {volume} {112}},\ \bibinfo {pages} {206402} (\bibinfo {year}
  {2014})}\BibitemShut {NoStop}%
\bibitem [{\citenamefont {Qiao}\ \emph {et~al.}(2017)\citenamefont {Qiao},
  \citenamefont {Li}, \citenamefont {Wang}, \citenamefont {Ruan}, \citenamefont
  {Ye}, \citenamefont {Cai}, \citenamefont {Hao}, \citenamefont {Yao},
  \citenamefont {Chen}, \citenamefont {Wu}, \citenamefont {Wang},\ and\
  \citenamefont {Liu}}]{qiao2017mottness}%
  \BibitemOpen
  \bibfield  {author} {\bibinfo {author} {\bibfnamefont {S.}~\bibnamefont
  {Qiao}}, \bibinfo {author} {\bibfnamefont {X.}~\bibnamefont {Li}}, \bibinfo
  {author} {\bibfnamefont {N.}~\bibnamefont {Wang}}, \bibinfo {author}
  {\bibfnamefont {W.}~\bibnamefont {Ruan}}, \bibinfo {author} {\bibfnamefont
  {C.}~\bibnamefont {Ye}}, \bibinfo {author} {\bibfnamefont {P.}~\bibnamefont
  {Cai}}, \bibinfo {author} {\bibfnamefont {Z.}~\bibnamefont {Hao}}, \bibinfo
  {author} {\bibfnamefont {H.}~\bibnamefont {Yao}}, \bibinfo {author}
  {\bibfnamefont {X.}~\bibnamefont {Chen}}, \bibinfo {author} {\bibfnamefont
  {J.}~\bibnamefont {Wu}}, \bibinfo {author} {\bibfnamefont {Y.}~\bibnamefont
  {Wang}},\ and\ \bibinfo {author} {\bibfnamefont {Z.}~\bibnamefont {Liu}},\
  }\bibfield  {title} {\bibinfo {title} {{Mottness Collapse in
  $1\mathrm{T}\text{\ensuremath{-}}{\mathrm{TaS}}_{2\ensuremath{-}x}{\mathrm{Se}}_{x}$
  Transition-Metal Dichalcogenide: An Interplay between Localized and Itinerant
  Orbitals}},\ }\href {https://doi.org/10.1103/PhysRevX.7.041054} {\bibfield
  {journal} {\bibinfo  {journal} {Phys. Rev. X}\ }\textbf {\bibinfo {volume}
  {7}},\ \bibinfo {pages} {041054} (\bibinfo {year} {2017})}\BibitemShut
  {NoStop}%
\bibitem [{\citenamefont {Yu}\ \emph {et~al.}(2017)\citenamefont {Yu},
  \citenamefont {Liu}, \citenamefont {Quan}, \citenamefont {Wu}, \citenamefont
  {Lin}, \citenamefont {Chang},\ and\ \citenamefont {Zou}}]{yu2017electronic}%
  \BibitemOpen
  \bibfield  {author} {\bibinfo {author} {\bibfnamefont {X.-L.}\ \bibnamefont
  {Yu}}, \bibinfo {author} {\bibfnamefont {D.-Y.}\ \bibnamefont {Liu}},
  \bibinfo {author} {\bibfnamefont {Y.-M.}\ \bibnamefont {Quan}}, \bibinfo
  {author} {\bibfnamefont {J.}~\bibnamefont {Wu}}, \bibinfo {author}
  {\bibfnamefont {H.-Q.}\ \bibnamefont {Lin}}, \bibinfo {author} {\bibfnamefont
  {K.}~\bibnamefont {Chang}},\ and\ \bibinfo {author} {\bibfnamefont {L.-J.}\
  \bibnamefont {Zou}},\ }\bibfield  {title} {\bibinfo {title} {{Electronic
  correlation effects and orbital density wave in the layered compound
  $1T{\text{-TaS}}_{2}$}},\ }\href {https://doi.org/10.1103/PhysRevB.96.125138}
  {\bibfield  {journal} {\bibinfo  {journal} {Phys. Rev. B}\ }\textbf {\bibinfo
  {volume} {96}},\ \bibinfo {pages} {125138} (\bibinfo {year}
  {2017})}\BibitemShut {NoStop}%
\bibitem [{\citenamefont {Darancet}\ \emph {et~al.}(2014)\citenamefont
  {Darancet}, \citenamefont {Millis},\ and\ \citenamefont
  {Marianetti}}]{darancet2014three}%
  \BibitemOpen
  \bibfield  {author} {\bibinfo {author} {\bibfnamefont {P.}~\bibnamefont
  {Darancet}}, \bibinfo {author} {\bibfnamefont {A.~J.}\ \bibnamefont
  {Millis}},\ and\ \bibinfo {author} {\bibfnamefont {C.~A.}\ \bibnamefont
  {Marianetti}},\ }\bibfield  {title} {\bibinfo {title} {Three-dimensional
  metallic and two-dimensional insulating behavior in octahedral tantalum
  dichalcogenides},\ }\href {https://doi.org/10.1103/PhysRevB.90.045134}
  {\bibfield  {journal} {\bibinfo  {journal} {Phys. Rev. B}\ }\textbf {\bibinfo
  {volume} {90}},\ \bibinfo {pages} {045134} (\bibinfo {year}
  {2014})}\BibitemShut {NoStop}%
\bibitem [{\citenamefont {Calandra}(2018)}]{calandra2018phonon}%
  \BibitemOpen
  \bibfield  {author} {\bibinfo {author} {\bibfnamefont {M.}~\bibnamefont
  {Calandra}},\ }\bibfield  {title} {\bibinfo {title} {{Phonon-Assisted
  Magnetic Mott-Insulating State in the Charge Density Wave Phase of
  Single-Layer $1T\text{\ensuremath{-}}{\mathrm{NbSe}}_{2}$}},\ }\href
  {https://doi.org/10.1103/PhysRevLett.121.026401} {\bibfield  {journal}
  {\bibinfo  {journal} {Phys. Rev. Lett.}\ }\textbf {\bibinfo {volume} {121}},\
  \bibinfo {pages} {026401} (\bibinfo {year} {2018})}\BibitemShut {NoStop}%
\bibitem [{\citenamefont {Kamil}\ \emph {et~al.}(2018)\citenamefont {Kamil},
  \citenamefont {Berges}, \citenamefont {Sch{\"o}nhoff}, \citenamefont
  {R{\"o}sner}, \citenamefont {Sch{\"u}ler}, \citenamefont {Sangiovanni},\ and\
  \citenamefont {Wehling}}]{kamil2018electronic}%
  \BibitemOpen
  \bibfield  {author} {\bibinfo {author} {\bibfnamefont {E.}~\bibnamefont
  {Kamil}}, \bibinfo {author} {\bibfnamefont {J.}~\bibnamefont {Berges}},
  \bibinfo {author} {\bibfnamefont {G.}~\bibnamefont {Sch{\"o}nhoff}}, \bibinfo
  {author} {\bibfnamefont {M.}~\bibnamefont {R{\"o}sner}}, \bibinfo {author}
  {\bibfnamefont {M.}~\bibnamefont {Sch{\"u}ler}}, \bibinfo {author}
  {\bibfnamefont {G.}~\bibnamefont {Sangiovanni}},\ and\ \bibinfo {author}
  {\bibfnamefont {T.}~\bibnamefont {Wehling}},\ }\bibfield  {title} {\bibinfo
  {title} {{Electronic structure of single layer 1T-NbSe2: interplay of lattice
  distortions, non-local exchange, and Mott--Hubbard correlations}},\ }\href
  {https://doi.org/10.1088/1361-648X/aad215} {\bibfield  {journal} {\bibinfo
  {journal} {Journal of Physics: Condensed Matter}\ }\textbf {\bibinfo {volume}
  {30}},\ \bibinfo {pages} {325601} (\bibinfo {year} {2018})}\BibitemShut
  {NoStop}%
\bibitem [{\citenamefont {Du}\ \emph {et~al.}(2024)\citenamefont {Du},
  \citenamefont {Jiang}, \citenamefont {Zheng}, \citenamefont {Zhang},
  \citenamefont {Wang},\ and\ \citenamefont {Zhang}}]{du2024theoretical}%
  \BibitemOpen
  \bibfield  {author} {\bibinfo {author} {\bibfnamefont {H.}~\bibnamefont
  {Du}}, \bibinfo {author} {\bibfnamefont {Z.}~\bibnamefont {Jiang}}, \bibinfo
  {author} {\bibfnamefont {J.}~\bibnamefont {Zheng}}, \bibinfo {author}
  {\bibfnamefont {X.}~\bibnamefont {Zhang}}, \bibinfo {author} {\bibfnamefont
  {W.}~\bibnamefont {Wang}},\ and\ \bibinfo {author} {\bibfnamefont
  {Z.}~\bibnamefont {Zhang}},\ }\bibfield  {title} {\bibinfo {title}
  {{Theoretical study of CDW phases for bulk NbX$_2$ (X= S and Se)}},\ }\href
  {https://doi.org/10.1039/D3CP04426B} {\bibfield  {journal} {\bibinfo
  {journal} {Physical Chemistry Chemical Physics}\ }\textbf {\bibinfo {volume}
  {26}},\ \bibinfo {pages} {2376} (\bibinfo {year} {2024})}\BibitemShut
  {NoStop}%
\bibitem [{\citenamefont {Guster}\ \emph {et~al.}(2019)\citenamefont {Guster},
  \citenamefont {Rubio-Verdu}, \citenamefont {Robles}, \citenamefont
  {Zaldívar}, \citenamefont {Dreher}, \citenamefont {Pruneda}, \citenamefont
  {Silva-Guillen}, \citenamefont {Choi}, \citenamefont {Pascual}, \citenamefont
  {Ugeda}, \citenamefont {Ordejon},\ and\ \citenamefont {Canadell}}]{Guster}%
  \BibitemOpen
  \bibfield  {author} {\bibinfo {author} {\bibfnamefont {B.}~\bibnamefont
  {Guster}}, \bibinfo {author} {\bibfnamefont {C.}~\bibnamefont {Rubio-Verdu}},
  \bibinfo {author} {\bibfnamefont {R.}~\bibnamefont {Robles}}, \bibinfo
  {author} {\bibfnamefont {J.}~\bibnamefont {Zaldívar}}, \bibinfo {author}
  {\bibfnamefont {P.}~\bibnamefont {Dreher}}, \bibinfo {author} {\bibfnamefont
  {M.}~\bibnamefont {Pruneda}}, \bibinfo {author} {\bibfnamefont
  {J.}~\bibnamefont {Silva-Guillen}}, \bibinfo {author} {\bibfnamefont {D.-J.}\
  \bibnamefont {Choi}}, \bibinfo {author} {\bibfnamefont {J.~I.}\ \bibnamefont
  {Pascual}}, \bibinfo {author} {\bibfnamefont {M.~M.}\ \bibnamefont {Ugeda}},
  \bibinfo {author} {\bibfnamefont {P.}~\bibnamefont {Ordejon}},\ and\ \bibinfo
  {author} {\bibfnamefont {E.}~\bibnamefont {Canadell}},\ }\bibfield  {title}
  {\bibinfo {title} {Coexistence of elastic modulations in the charge density
  wave state of {2H‑NbSe2}},\ }\href
  {https://doi.org/ 10.1021/acs.nanolett.9b00268} {\bibfield  {journal}
  {\bibinfo  {journal} {Nano Lett.}\ }\textbf {\bibinfo {volume} { 19}},\
  \bibinfo {pages} {3027–3032} (\bibinfo {year} {2019})}\BibitemShut
  {NoStop}%
\bibitem [{\citenamefont {Malliakas}\ and\ \citenamefont
  {Kanatzidis}(2013)}]{Malliakas}%
  \BibitemOpen
  \bibfield  {author} {\bibinfo {author} {\bibfnamefont {C.~D.}\ \bibnamefont
  {Malliakas}}\ and\ \bibinfo {author} {\bibfnamefont {M.~G.}\ \bibnamefont
  {Kanatzidis}},\ }\bibfield  {title} {\bibinfo {title} {Nb-nb interactions
  define the charge density wave structure of {2H-NbSe2}},\ }\href
  {https://doi.org/10.1021/ja3120554} {\bibfield  {journal} {\bibinfo
  {journal} {J. Am. Chem. Soc.}\ }\textbf {\bibinfo {volume} { 135}},\
  \bibinfo {pages} {1719–1722} (\bibinfo {year} {2013})}\BibitemShut
  {NoStop}%
\bibitem [{\citenamefont {Calandra}\ \emph {et~al.}(2009)\citenamefont
  {Calandra}, \citenamefont {Mazin},\ and\ \citenamefont {Mauri}}]{Calandra}%
  \BibitemOpen
  \bibfield  {author} {\bibinfo {author} {\bibfnamefont {M.}~\bibnamefont
  {Calandra}}, \bibinfo {author} {\bibfnamefont {I.~I.}\ \bibnamefont
  {Mazin}},\ and\ \bibinfo {author} {\bibfnamefont {F.}~\bibnamefont {Mauri}},\
  }\bibfield  {title} {\bibinfo {title} {Effect of dimensionality on the
  charge-density wave in few-layer {2H-NbSe2}},\ }\href
  {https://doi.org/10.1103/PhysRevB.80.241108} {\bibfield  {journal} {\bibinfo
  {journal} {Phys. Rev. B}\ }\textbf {\bibinfo {volume} { 80}},\ \bibinfo
  {pages} {241108} (\bibinfo {year} {2009})}\BibitemShut {NoStop}%
\bibitem [{\citenamefont {Arguello}\ \emph {et~al.}(2014)\citenamefont
  {Arguello}, \citenamefont {Chockalingam}, \citenamefont {Rosenthal},
  \citenamefont {Zhao}, \citenamefont {Gutierrez}, \citenamefont {Kang},
  \citenamefont {Chung}, \citenamefont {Fernandes}, \citenamefont {Jia},
  \citenamefont {Millis}, \citenamefont {Cava},\ and\ \citenamefont
  {Pasupathy}}]{Arguello}%
  \BibitemOpen
  \bibfield  {author} {\bibinfo {author} {\bibfnamefont {C.~J.}\ \bibnamefont
  {Arguello}}, \bibinfo {author} {\bibfnamefont {S.~P.}\ \bibnamefont
  {Chockalingam}}, \bibinfo {author} {\bibfnamefont {E.~P.}\ \bibnamefont
  {Rosenthal}}, \bibinfo {author} {\bibfnamefont {L.}~\bibnamefont {Zhao}},
  \bibinfo {author} {\bibfnamefont {C.}~\bibnamefont {Gutierrez}}, \bibinfo
  {author} {\bibfnamefont {J.~H.}\ \bibnamefont {Kang}}, \bibinfo {author}
  {\bibfnamefont {W.~C.}\ \bibnamefont {Chung}}, \bibinfo {author}
  {\bibfnamefont {R.~M.}\ \bibnamefont {Fernandes}}, \bibinfo {author}
  {\bibfnamefont {S.}~\bibnamefont {Jia}}, \bibinfo {author} {\bibfnamefont
  {A.~J.}\ \bibnamefont {Millis}}, \bibinfo {author} {\bibfnamefont {R.~J.}\
  \bibnamefont {Cava}},\ and\ \bibinfo {author} {\bibfnamefont {A.~N.}\
  \bibnamefont {Pasupathy}},\ }\bibfield  {title} {\bibinfo {title}
  {Visualizing the charge density wave transition in {2H-NbSe2} in real
  space},\ }\href {https://doi.org/ 10.1103/PhysRevB.89.235115} {\bibfield
  {journal} {\bibinfo  {journal} {Phys. Rev. B}\ }\textbf {\bibinfo {volume}
  {89}},\ \bibinfo {pages} {235115} (\bibinfo {year} {2014})}\BibitemShut
  {NoStop}%
\bibitem [{\citenamefont {Straub}\ \emph
  {et~al.}(1999{\natexlab{a}})\citenamefont {Straub}, \citenamefont {Finteis},
  \citenamefont {Claessen}, \citenamefont {Steiner}, \citenamefont {Hüfner},
  \citenamefont {Blaha}, \citenamefont {Oglesby},\ and\ \citenamefont
  {Bucher}}]{Straub}%
  \BibitemOpen
  \bibfield  {author} {\bibinfo {author} {\bibfnamefont {T.}~\bibnamefont
  {Straub}}, \bibinfo {author} {\bibfnamefont {T.}~\bibnamefont {Finteis}},
  \bibinfo {author} {\bibfnamefont {R.}~\bibnamefont {Claessen}}, \bibinfo
  {author} {\bibfnamefont {P.}~\bibnamefont {Steiner}}, \bibinfo {author}
  {\bibfnamefont {S.}~\bibnamefont {Hüfner}}, \bibinfo {author} {\bibfnamefont
  {P.}~\bibnamefont {Blaha}}, \bibinfo {author} {\bibfnamefont {C.~S.}\
  \bibnamefont {Oglesby}},\ and\ \bibinfo {author} {\bibfnamefont
  {E.}~\bibnamefont {Bucher}},\ }\bibfield  {title} {\bibinfo {title}
  {Charge-density-wave mechanism in {2H-NbSe2:} photoemission results},\
  }\href {https://doi.org/10.1103/PhysRevLett.82.4504} {\bibfield  {journal}
  {\bibinfo  {journal} {Phys. Rev. Lett}\ }\textbf {\bibinfo {volume} {82}},\
  \bibinfo {pages} {4504} (\bibinfo {year} {1999}{\natexlab{a}})}\BibitemShut
  {NoStop}%
\bibitem [{\citenamefont {Wang}\ \emph {et~al.}(2017)\citenamefont {Wang},
  \citenamefont {Huang}, \citenamefont {Lin}, \citenamefont {Cui},
  \citenamefont {Chen}, \citenamefont {Zhu}, \citenamefont {Liu}, \citenamefont
  {Zeng}, \citenamefont {Zhou}, \citenamefont {Yu} \emph
  {et~al.}}]{wang2017high}%
  \BibitemOpen
  \bibfield  {author} {\bibinfo {author} {\bibfnamefont {H.}~\bibnamefont
  {Wang}}, \bibinfo {author} {\bibfnamefont {X.}~\bibnamefont {Huang}},
  \bibinfo {author} {\bibfnamefont {J.}~\bibnamefont {Lin}}, \bibinfo {author}
  {\bibfnamefont {J.}~\bibnamefont {Cui}}, \bibinfo {author} {\bibfnamefont
  {Y.}~\bibnamefont {Chen}}, \bibinfo {author} {\bibfnamefont {C.}~\bibnamefont
  {Zhu}}, \bibinfo {author} {\bibfnamefont {F.}~\bibnamefont {Liu}}, \bibinfo
  {author} {\bibfnamefont {Q.}~\bibnamefont {Zeng}}, \bibinfo {author}
  {\bibfnamefont {J.}~\bibnamefont {Zhou}}, \bibinfo {author} {\bibfnamefont
  {P.}~\bibnamefont {Yu}}, \emph {et~al.},\ }\bibfield  {title} {\bibinfo
  {title} {High-quality monolayer superconductor nbse2 grown by chemical vapour
  deposition},\ }\href {https://doi.org/10.1038/s41467-017-00427-5} {\bibfield
  {journal} {\bibinfo  {journal} {Nature communications}\ }\textbf {\bibinfo
  {volume} {8}},\ \bibinfo {pages} {394} (\bibinfo {year} {2017})}\BibitemShut
  {NoStop}%
\bibitem [{\citenamefont {Hamill}\ \emph {et~al.}(2021)\citenamefont {Hamill},
  \citenamefont {Heischmidt}, \citenamefont {Sohn}, \citenamefont {Shaffer},
  \citenamefont {Tsai}, \citenamefont {Zhang}, \citenamefont {Xi},
  \citenamefont {Suslov}, \citenamefont {Berger}, \citenamefont {Forr{\'o}}
  \emph {et~al.}}]{hamill2021two}%
  \BibitemOpen
  \bibfield  {author} {\bibinfo {author} {\bibfnamefont {A.}~\bibnamefont
  {Hamill}}, \bibinfo {author} {\bibfnamefont {B.}~\bibnamefont {Heischmidt}},
  \bibinfo {author} {\bibfnamefont {E.}~\bibnamefont {Sohn}}, \bibinfo {author}
  {\bibfnamefont {D.}~\bibnamefont {Shaffer}}, \bibinfo {author} {\bibfnamefont
  {K.-T.}\ \bibnamefont {Tsai}}, \bibinfo {author} {\bibfnamefont
  {X.}~\bibnamefont {Zhang}}, \bibinfo {author} {\bibfnamefont
  {X.}~\bibnamefont {Xi}}, \bibinfo {author} {\bibfnamefont {A.}~\bibnamefont
  {Suslov}}, \bibinfo {author} {\bibfnamefont {H.}~\bibnamefont {Berger}},
  \bibinfo {author} {\bibfnamefont {L.}~\bibnamefont {Forr{\'o}}}, \emph
  {et~al.},\ }\bibfield  {title} {\bibinfo {title} {{Two-fold symmetric
  superconductivity in few-layer NbSe$_2$}},\ }\href
  {https://doi.org/10.1038/s41567-021-01219-x} {\bibfield  {journal} {\bibinfo
  {journal} {Nature physics}\ }\textbf {\bibinfo {volume} {17}},\ \bibinfo
  {pages} {949} (\bibinfo {year} {2021})}\BibitemShut {NoStop}%
\bibitem [{\citenamefont {Patel}\ \emph {et~al.}(2024)\citenamefont {Patel},
  \citenamefont {Jena},\ and\ \citenamefont {Taraphder}}]{patel2024electron}%
  \BibitemOpen
  \bibfield  {author} {\bibinfo {author} {\bibfnamefont {S.}~\bibnamefont
  {Patel}}, \bibinfo {author} {\bibfnamefont {S.}~\bibnamefont {Jena}},\ and\
  \bibinfo {author} {\bibfnamefont {A.}~\bibnamefont {Taraphder}},\ }\bibfield
  {title} {\bibinfo {title} {{Electron-phonon coupling, critical temperatures,
  and gaps in ${\mathrm{NbSe}}_{2}/{\mathrm{MoS}}_{2}$ Ising
  superconductors}},\ }\href {https://doi.org/10.1103/PhysRevB.110.014507}
  {\bibfield  {journal} {\bibinfo  {journal} {Phys. Rev. B}\ }\textbf {\bibinfo
  {volume} {110}},\ \bibinfo {pages} {014507} (\bibinfo {year}
  {2024})}\BibitemShut {NoStop}%
\bibitem [{\citenamefont {Xi}\ \emph {et~al.}(2016)\citenamefont {Xi},
  \citenamefont {Wang}, \citenamefont {Zhao}, \citenamefont {Park},
  \citenamefont {Law}, \citenamefont {Berger}, \citenamefont {Forr{\'o}},
  \citenamefont {Shan},\ and\ \citenamefont {Mak}}]{xi2016ising}%
  \BibitemOpen
  \bibfield  {author} {\bibinfo {author} {\bibfnamefont {X.}~\bibnamefont
  {Xi}}, \bibinfo {author} {\bibfnamefont {Z.}~\bibnamefont {Wang}}, \bibinfo
  {author} {\bibfnamefont {W.}~\bibnamefont {Zhao}}, \bibinfo {author}
  {\bibfnamefont {J.-H.}\ \bibnamefont {Park}}, \bibinfo {author}
  {\bibfnamefont {K.~T.}\ \bibnamefont {Law}}, \bibinfo {author} {\bibfnamefont
  {H.}~\bibnamefont {Berger}}, \bibinfo {author} {\bibfnamefont
  {L.}~\bibnamefont {Forr{\'o}}}, \bibinfo {author} {\bibfnamefont
  {J.}~\bibnamefont {Shan}},\ and\ \bibinfo {author} {\bibfnamefont {K.~F.}\
  \bibnamefont {Mak}},\ }\bibfield  {title} {\bibinfo {title} {{Ising pairing
  in superconducting NbSe$_2$ atomic layers}},\ }\href
  {https://doi.org/10.1038/nphys3538} {\bibfield  {journal} {\bibinfo
  {journal} {Nature Physics}\ }\textbf {\bibinfo {volume} {12}},\ \bibinfo
  {pages} {139} (\bibinfo {year} {2016})}\BibitemShut {NoStop}%
\bibitem [{\citenamefont {Wickramaratne}\ \emph {et~al.}(2020)\citenamefont
  {Wickramaratne}, \citenamefont {Khmelevskyi}, \citenamefont {Agterberg},\
  and\ \citenamefont {Mazin}}]{wickramaratne2020ising}%
  \BibitemOpen
  \bibfield  {author} {\bibinfo {author} {\bibfnamefont {D.}~\bibnamefont
  {Wickramaratne}}, \bibinfo {author} {\bibfnamefont {S.}~\bibnamefont
  {Khmelevskyi}}, \bibinfo {author} {\bibfnamefont {D.~F.}\ \bibnamefont
  {Agterberg}},\ and\ \bibinfo {author} {\bibfnamefont {I.~I.}\ \bibnamefont
  {Mazin}},\ }\bibfield  {title} {\bibinfo {title} {Ising superconductivity and
  magnetism in ${\mathrm{nbse}}_{2}$},\ }\href
  {https://doi.org/10.1103/PhysRevX.10.041003} {\bibfield  {journal} {\bibinfo
  {journal} {Phys. Rev. X}\ }\textbf {\bibinfo {volume} {10}},\ \bibinfo
  {pages} {041003} (\bibinfo {year} {2020})}\BibitemShut {NoStop}%
\bibitem [{\citenamefont {Das}\ \emph {et~al.}(2023)\citenamefont {Das},
  \citenamefont {Paudyal}, \citenamefont {Margine}, \citenamefont {Agterberg},\
  and\ \citenamefont {Mazin}}]{das2023electron}%
  \BibitemOpen
  \bibfield  {author} {\bibinfo {author} {\bibfnamefont {S.}~\bibnamefont
  {Das}}, \bibinfo {author} {\bibfnamefont {H.}~\bibnamefont {Paudyal}},
  \bibinfo {author} {\bibfnamefont {E.}~\bibnamefont {Margine}}, \bibinfo
  {author} {\bibfnamefont {D.}~\bibnamefont {Agterberg}},\ and\ \bibinfo
  {author} {\bibfnamefont {I.}~\bibnamefont {Mazin}},\ }\bibfield  {title}
  {\bibinfo {title} {Electron-phonon coupling and spin fluctuations in the
  ising superconductor nbse2},\ }\href
  {https://doi.org/10.1038/s41524-023-01017-4} {\bibfield  {journal} {\bibinfo
  {journal} {npj Computational Materials}\ }\textbf {\bibinfo {volume} {9}},\
  \bibinfo {pages} {66} (\bibinfo {year} {2023})}\BibitemShut {NoStop}%
\bibitem [{\citenamefont {Weber}\ \emph {et~al.}(2011)\citenamefont {Weber},
  \citenamefont {Rosenkranz}, \citenamefont {Castellan}, \citenamefont
  {Osborn}, \citenamefont {Hott}, \citenamefont {Heid}, \citenamefont {Bohnen},
  \citenamefont {Egami}, \citenamefont {Said},\ and\ \citenamefont
  {Reznik}}]{weber2011extended}%
  \BibitemOpen
  \bibfield  {author} {\bibinfo {author} {\bibfnamefont {F.}~\bibnamefont
  {Weber}}, \bibinfo {author} {\bibfnamefont {S.}~\bibnamefont {Rosenkranz}},
  \bibinfo {author} {\bibfnamefont {J.-P.}\ \bibnamefont {Castellan}}, \bibinfo
  {author} {\bibfnamefont {R.}~\bibnamefont {Osborn}}, \bibinfo {author}
  {\bibfnamefont {R.}~\bibnamefont {Hott}}, \bibinfo {author} {\bibfnamefont
  {R.}~\bibnamefont {Heid}}, \bibinfo {author} {\bibfnamefont {K.-P.}\
  \bibnamefont {Bohnen}}, \bibinfo {author} {\bibfnamefont {T.}~\bibnamefont
  {Egami}}, \bibinfo {author} {\bibfnamefont {A.~H.}\ \bibnamefont {Said}},\
  and\ \bibinfo {author} {\bibfnamefont {D.}~\bibnamefont {Reznik}},\
  }\bibfield  {title} {\bibinfo {title} {{Extended Phonon Collapse and the
  Origin of the Charge-Density Wave in
  $2H\mathrm{\text{\ensuremath{-}}}{\mathrm{NbSe}}_{2}$}},\ }\href
  {https://doi.org/10.1103/PhysRevLett.107.107403} {\bibfield  {journal}
  {\bibinfo  {journal} {Phys. Rev. Lett.}\ }\textbf {\bibinfo {volume} {107}},\
  \bibinfo {pages} {107403} (\bibinfo {year} {2011})}\BibitemShut {NoStop}%
\bibitem [{\citenamefont {Ugeda}\ \emph {et~al.}(2016)\citenamefont {Ugeda},
  \citenamefont {Bradley}, \citenamefont {Zhang}, \citenamefont {Onishi},
  \citenamefont {Chen}, \citenamefont {Ruan}, \citenamefont
  {Ojeda-Aristizabal}, \citenamefont {Ryu}, \citenamefont {Edmonds},
  \citenamefont {Tsai} \emph {et~al.}}]{ugeda2016characterization}%
  \BibitemOpen
  \bibfield  {author} {\bibinfo {author} {\bibfnamefont {M.~M.}\ \bibnamefont
  {Ugeda}}, \bibinfo {author} {\bibfnamefont {A.~J.}\ \bibnamefont {Bradley}},
  \bibinfo {author} {\bibfnamefont {Y.}~\bibnamefont {Zhang}}, \bibinfo
  {author} {\bibfnamefont {S.}~\bibnamefont {Onishi}}, \bibinfo {author}
  {\bibfnamefont {Y.}~\bibnamefont {Chen}}, \bibinfo {author} {\bibfnamefont
  {W.}~\bibnamefont {Ruan}}, \bibinfo {author} {\bibfnamefont {C.}~\bibnamefont
  {Ojeda-Aristizabal}}, \bibinfo {author} {\bibfnamefont {H.}~\bibnamefont
  {Ryu}}, \bibinfo {author} {\bibfnamefont {M.~T.}\ \bibnamefont {Edmonds}},
  \bibinfo {author} {\bibfnamefont {H.-Z.}\ \bibnamefont {Tsai}}, \emph
  {et~al.},\ }\bibfield  {title} {\bibinfo {title} {Characterization of
  collective ground states in single-layer nbse 2},\ }\href
  {https://doi.org/10.1038/nphys3527} {\bibfield  {journal} {\bibinfo
  {journal} {Nature Physics}\ }\textbf {\bibinfo {volume} {12}},\ \bibinfo
  {pages} {92} (\bibinfo {year} {2016})}\BibitemShut {NoStop}%
\bibitem [{\citenamefont {Koley}\ \emph {et~al.}(2020)\citenamefont {Koley},
  \citenamefont {Mohanta},\ and\ \citenamefont {Taraphder}}]{koley2020charge}%
  \BibitemOpen
  \bibfield  {author} {\bibinfo {author} {\bibfnamefont {S.}~\bibnamefont
  {Koley}}, \bibinfo {author} {\bibfnamefont {N.}~\bibnamefont {Mohanta}},\
  and\ \bibinfo {author} {\bibfnamefont {A.}~\bibnamefont {Taraphder}},\
  }\bibfield  {title} {\bibinfo {title} {Charge density wave and
  superconductivity in transition metal dichalcogenides},\ }\href
  {https://doi.org/10.1140/epjb/e2020-100522-5} {\bibfield  {journal} {\bibinfo
   {journal} {The European Physical Journal B}\ }\textbf {\bibinfo {volume}
  {93}},\ \bibinfo {pages} {1} (\bibinfo {year} {2020})}\BibitemShut {NoStop}%
\bibitem [{\citenamefont {Feng}\ \emph {et~al.}(2023)\citenamefont {Feng},
  \citenamefont {Cao}, \citenamefont {Priessnitz}, \citenamefont {Dai},
  \citenamefont {de~Oliveira}, \citenamefont {Yuan}, \citenamefont {Oka},
  \citenamefont {Kim}, \citenamefont {Chen}, \citenamefont {Ponomaryov},
  \citenamefont {Ilyakov}, \citenamefont {Zhang}, \citenamefont {Lv},
  \citenamefont {Mazzotti}, \citenamefont {Kim}, \citenamefont {Christiani},
  \citenamefont {Logvenov}, \citenamefont {Wu}, \citenamefont {Huang},
  \citenamefont {Deinert}, \citenamefont {Kovalev}, \citenamefont {Kaiser},
  \citenamefont {Dong}, \citenamefont {Wang},\ and\ \citenamefont
  {Chu}}]{feng2023dynamical}%
  \BibitemOpen
  \bibfield  {author} {\bibinfo {author} {\bibfnamefont {L.}~\bibnamefont
  {Feng}}, \bibinfo {author} {\bibfnamefont {J.}~\bibnamefont {Cao}}, \bibinfo
  {author} {\bibfnamefont {T.}~\bibnamefont {Priessnitz}}, \bibinfo {author}
  {\bibfnamefont {Y.}~\bibnamefont {Dai}}, \bibinfo {author} {\bibfnamefont
  {T.}~\bibnamefont {de~Oliveira}}, \bibinfo {author} {\bibfnamefont
  {J.}~\bibnamefont {Yuan}}, \bibinfo {author} {\bibfnamefont {R.}~\bibnamefont
  {Oka}}, \bibinfo {author} {\bibfnamefont {M.-J.}\ \bibnamefont {Kim}},
  \bibinfo {author} {\bibfnamefont {M.}~\bibnamefont {Chen}}, \bibinfo {author}
  {\bibfnamefont {A.~N.}\ \bibnamefont {Ponomaryov}}, \bibinfo {author}
  {\bibfnamefont {I.}~\bibnamefont {Ilyakov}}, \bibinfo {author} {\bibfnamefont
  {H.}~\bibnamefont {Zhang}}, \bibinfo {author} {\bibfnamefont
  {Y.}~\bibnamefont {Lv}}, \bibinfo {author} {\bibfnamefont {V.}~\bibnamefont
  {Mazzotti}}, \bibinfo {author} {\bibfnamefont {G.}~\bibnamefont {Kim}},
  \bibinfo {author} {\bibfnamefont {G.}~\bibnamefont {Christiani}}, \bibinfo
  {author} {\bibfnamefont {G.}~\bibnamefont {Logvenov}}, \bibinfo {author}
  {\bibfnamefont {D.}~\bibnamefont {Wu}}, \bibinfo {author} {\bibfnamefont
  {Y.}~\bibnamefont {Huang}}, \bibinfo {author} {\bibfnamefont {J.-C.}\
  \bibnamefont {Deinert}}, \bibinfo {author} {\bibfnamefont {S.}~\bibnamefont
  {Kovalev}}, \bibinfo {author} {\bibfnamefont {S.}~\bibnamefont {Kaiser}},
  \bibinfo {author} {\bibfnamefont {T.}~\bibnamefont {Dong}}, \bibinfo {author}
  {\bibfnamefont {N.}~\bibnamefont {Wang}},\ and\ \bibinfo {author}
  {\bibfnamefont {H.}~\bibnamefont {Chu}},\ }\bibfield  {title} {\bibinfo
  {title} {{Dynamical interplay between superconductivity and charge density
  waves: A nonlinear terahertz study of coherently driven
  $2H\text{\ensuremath{-}}{\mathrm{NbSe}}_{2}$}},\ }\href
  {https://doi.org/10.1103/PhysRevB.108.L100504} {\bibfield  {journal}
  {\bibinfo  {journal} {Phys. Rev. B}\ }\textbf {\bibinfo {volume} {108}},\
  \bibinfo {pages} {L100504} (\bibinfo {year} {2023})}\BibitemShut {NoStop}%
\bibitem [{\citenamefont {Burk}\ \emph {et~al.}(1991)\citenamefont {Burk},
  \citenamefont {Thomson}, \citenamefont {Zettl},\ and\ \citenamefont
  {Clarke}}]{Burk}%
  \BibitemOpen
  \bibfield  {author} {\bibinfo {author} {\bibfnamefont {B.}~\bibnamefont
  {Burk}}, \bibinfo {author} {\bibfnamefont {R.~E.}\ \bibnamefont {Thomson}},
  \bibinfo {author} {\bibfnamefont {A.}~\bibnamefont {Zettl}},\ and\ \bibinfo
  {author} {\bibfnamefont {J.}~\bibnamefont {Clarke}},\ }\bibfield  {title}
  {\bibinfo {title} {Charge-density-wave domains in 1t-tas2 observed by
  satellite structure in scanning-tunneling-microscopy images},\ }\href
  {https://doi.org/10.1103/PhysRevLett.66.3040} {\bibfield  {journal} {\bibinfo
   {journal} {Phys. Rev. Lett. }\ }\textbf {\bibinfo {volume} {66}},\ \bibinfo
  {pages} {3040} (\bibinfo {year} {1991})}\BibitemShut {NoStop}%
\bibitem [{\citenamefont {and Robert Hovden and Dennis Wang}\ and\
  \citenamefont {N. Pasupathy }(2015)}]{Tsen}%
  \BibitemOpen
  \bibfield  {author} {\bibinfo {author} {\bibfnamefont {A.~W.}\ \bibnamefont
  {and Robert Hovden and Dennis Wang}}\ and\ \bibinfo {author}
  {\bibfnamefont {A.}~\bibnamefont {N. Pasupathy }},\ }\bibfield  {title}
  {\bibinfo {title} {Structure and control of charge density waves in
  two-dimensional {1T-TaS2}},\ }\href {https://doi.org/10.1073/pnas.1512092112}
  {\bibfield  {journal} {\bibinfo  {journal} {PNAS}\ }\textbf {\bibinfo
  {volume} {112}},\ \bibinfo {pages} {15054} (\bibinfo {year}
  {2015})}\BibitemShut {NoStop}%
\bibitem [{\citenamefont {Freitas}\ \emph {et~al.}(2016)\citenamefont
  {Freitas}, \citenamefont {Rodi{\`e}re}, \citenamefont {Osorio}, \citenamefont
  {Navarro-Moratalla}, \citenamefont {Nemes}, \citenamefont {Tissen},
  \citenamefont {Cario}, \citenamefont {Coronado}, \citenamefont
  {Garc{\'{\i}}a-Hern{\'a}ndez}, \citenamefont {Vieira}, \citenamefont
  {N{\'u}{\~n}ez-Regueiro},\ and\ \citenamefont {Suderow}}]{freitas2016strong}%
  \BibitemOpen
  \bibfield  {author} {\bibinfo {author} {\bibfnamefont {D.~C.}\ \bibnamefont
  {Freitas}}, \bibinfo {author} {\bibfnamefont {P.}~\bibnamefont
  {Rodi{\`e}re}}, \bibinfo {author} {\bibfnamefont {M.~R.}\ \bibnamefont
  {Osorio}}, \bibinfo {author} {\bibfnamefont {E.}~\bibnamefont
  {Navarro-Moratalla}}, \bibinfo {author} {\bibfnamefont {N.~M.}\ \bibnamefont
  {Nemes}}, \bibinfo {author} {\bibfnamefont {V.~G.}\ \bibnamefont {Tissen}},
  \bibinfo {author} {\bibfnamefont {L.}~\bibnamefont {Cario}}, \bibinfo
  {author} {\bibfnamefont {E.}~\bibnamefont {Coronado}}, \bibinfo {author}
  {\bibfnamefont {M.}~\bibnamefont {Garc{\'{\i}}a-Hern{\'a}ndez}}, \bibinfo
  {author} {\bibfnamefont {S.}~\bibnamefont {Vieira}}, \bibinfo {author}
  {\bibfnamefont {M.}~\bibnamefont {N{\'u}{\~n}ez-Regueiro}},\ and\ \bibinfo
  {author} {\bibfnamefont {H.}~\bibnamefont {Suderow}},\ }\bibfield  {title}
  {\bibinfo {title} {Strong enhancement of superconductivity at high pressures
  within the charge-density-wave states of {2H-TaS$_2$ and 2H-TaSe$_2$}},\
  }\href {https://doi.org/10.1103/PhysRevB.93.184512} {\bibfield  {journal}
  {\bibinfo  {journal} {Phys. Rev. B}\ }\textbf {\bibinfo {volume} {93}},\
  \bibinfo {pages} {184512} (\bibinfo {year} {2016})}\BibitemShut {NoStop}%
\bibitem [{\citenamefont {Wang}\ \emph {et~al.}(2020)\citenamefont {Wang},
  \citenamefont {Yao}, \citenamefont {Xin}, \citenamefont {Han}, \citenamefont
  {Wang}, \citenamefont {Chen}, \citenamefont {Cai}, \citenamefont {Li},\ and\
  \citenamefont {Zhang}}]{wang2020band}%
  \BibitemOpen
  \bibfield  {author} {\bibinfo {author} {\bibfnamefont {Y.}~\bibnamefont
  {Wang}}, \bibinfo {author} {\bibfnamefont {W.}~\bibnamefont {Yao}}, \bibinfo
  {author} {\bibfnamefont {Z.}~\bibnamefont {Xin}}, \bibinfo {author}
  {\bibfnamefont {T.}~\bibnamefont {Han}}, \bibinfo {author} {\bibfnamefont
  {Z.}~\bibnamefont {Wang}}, \bibinfo {author} {\bibfnamefont {L.}~\bibnamefont
  {Chen}}, \bibinfo {author} {\bibfnamefont {C.}~\bibnamefont {Cai}}, \bibinfo
  {author} {\bibfnamefont {Y.}~\bibnamefont {Li}},\ and\ \bibinfo {author}
  {\bibfnamefont {Y.}~\bibnamefont {Zhang}},\ }\bibfield  {title} {\bibinfo
  {title} {Band insulator to mott insulator transition in 1 t-tas2},\ }\href
  {https://doi.org/10.1038/s41467-020-18040-4} {\bibfield  {journal} {\bibinfo
  {journal} {Nature communications}\ }\textbf {\bibinfo {volume} {11}},\
  \bibinfo {pages} {4215} (\bibinfo {year} {2020})}\BibitemShut {NoStop}%
\bibitem [{\citenamefont {J.A.~Wilson}\ and\ \citenamefont
  {Mahajan}(1975)}]{wilson1975charge}%
  \BibitemOpen
  \bibfield  {author} {\bibinfo {author} {\bibfnamefont {F.~D.~S.}\
  \bibnamefont {J.A.~Wilson}}\ and\ \bibinfo {author} {\bibfnamefont
  {S.}~\bibnamefont {Mahajan}},\ }\bibfield  {title} {\bibinfo {title}
  {Charge-density waves and superlattices in the metallic layered transition
  metal dichalcogenides},\ }\href {https://doi.org/10.1080/00018737500101391}
  {\bibfield  {journal} {\bibinfo  {journal} {Advances in Physics}\ }\textbf
  {\bibinfo {volume} {24}},\ \bibinfo {pages} {117} (\bibinfo {year}
  {1975})}\BibitemShut {NoStop}%
\bibitem [{\citenamefont {Rice}\ and\ \citenamefont
  {Scott}(1975)}]{rice1975new}%
  \BibitemOpen
  \bibfield  {author} {\bibinfo {author} {\bibfnamefont {T.~M.}\ \bibnamefont
  {Rice}}\ and\ \bibinfo {author} {\bibfnamefont {G.~K.}\ \bibnamefont
  {Scott}},\ }\bibfield  {title} {\bibinfo {title} {New mechanism for a
  charge-density-wave instability},\ }\href
  {https://doi.org/10.1103/PhysRevLett.35.120} {\bibfield  {journal} {\bibinfo
  {journal} {Phys. Rev. Lett.}\ }\textbf {\bibinfo {volume} {35}},\ \bibinfo
  {pages} {120} (\bibinfo {year} {1975})}\BibitemShut {NoStop}%
\bibitem [{\citenamefont {Johannes}\ \emph {et~al.}(2006)\citenamefont
  {Johannes}, \citenamefont {Mazin},\ and\ \citenamefont
  {Howells}}]{johannes2006fermi}%
  \BibitemOpen
  \bibfield  {author} {\bibinfo {author} {\bibfnamefont {M.~D.}\ \bibnamefont
  {Johannes}}, \bibinfo {author} {\bibfnamefont {I.~I.}\ \bibnamefont
  {Mazin}},\ and\ \bibinfo {author} {\bibfnamefont {C.~A.}\ \bibnamefont
  {Howells}},\ }\bibfield  {title} {\bibinfo {title} {Fermi-surface nesting and
  the origin of the charge-density wave in $\mathrm{Nb}{\mathrm{se}}_{2}$},\
  }\href {https://doi.org/10.1103/PhysRevB.73.205102} {\bibfield  {journal}
  {\bibinfo  {journal} {Phys. Rev. B}\ }\textbf {\bibinfo {volume} {73}},\
  \bibinfo {pages} {205102} (\bibinfo {year} {2006})}\BibitemShut {NoStop}%
\bibitem [{\citenamefont {Liu}\ \emph {et~al.}(1998)\citenamefont {Liu},
  \citenamefont {Olson}, \citenamefont {Tonjes},\ and\ \citenamefont
  {Frindt}}]{liu1998momentum}%
  \BibitemOpen
  \bibfield  {author} {\bibinfo {author} {\bibfnamefont {R.}~\bibnamefont
  {Liu}}, \bibinfo {author} {\bibfnamefont {C.~G.}\ \bibnamefont {Olson}},
  \bibinfo {author} {\bibfnamefont {W.~C.}\ \bibnamefont {Tonjes}},\ and\
  \bibinfo {author} {\bibfnamefont {R.~F.}\ \bibnamefont {Frindt}},\ }\bibfield
   {title} {\bibinfo {title} {{Momentum Dependent Spectral Changes Induced by
  the Charge Density Wave in $2\mathit{H}\ensuremath{-}{\mathrm{TaSe}}_{2}$ and
  the Implication on the CDW Mechanism}},\ }\href
  {https://doi.org/10.1103/PhysRevLett.80.5762} {\bibfield  {journal} {\bibinfo
   {journal} {Phys. Rev. Lett.}\ }\textbf {\bibinfo {volume} {80}},\ \bibinfo
  {pages} {5762} (\bibinfo {year} {1998})}\BibitemShut {NoStop}%
\bibitem [{\citenamefont {Liu}\ \emph {et~al.}(2000)\citenamefont {Liu},
  \citenamefont {Tonjes}, \citenamefont {Greanya}, \citenamefont {Olson},\ and\
  \citenamefont {Frindt}}]{liu2000fermi}%
  \BibitemOpen
  \bibfield  {author} {\bibinfo {author} {\bibfnamefont {R.}~\bibnamefont
  {Liu}}, \bibinfo {author} {\bibfnamefont {W.~C.}\ \bibnamefont {Tonjes}},
  \bibinfo {author} {\bibfnamefont {V.~A.}\ \bibnamefont {Greanya}}, \bibinfo
  {author} {\bibfnamefont {C.~G.}\ \bibnamefont {Olson}},\ and\ \bibinfo
  {author} {\bibfnamefont {R.~F.}\ \bibnamefont {Frindt}},\ }\bibfield  {title}
  {\bibinfo {title} {{Fermi surface of $2H\ensuremath{-}{\mathrm{TaSe}}_{2}$
  and its relation to the charge-density wave}},\ }\href
  {https://doi.org/10.1103/PhysRevB.61.5212} {\bibfield  {journal} {\bibinfo
  {journal} {Phys. Rev. B}\ }\textbf {\bibinfo {volume} {61}},\ \bibinfo
  {pages} {5212} (\bibinfo {year} {2000})}\BibitemShut {NoStop}%
\bibitem [{\citenamefont {Dardel}\ \emph {et~al.}(1993)\citenamefont {Dardel},
  \citenamefont {Grioni}, \citenamefont {Malterre}, \citenamefont {Weibel},
  \citenamefont {Baer},\ and\ \citenamefont {Levy}}]{dardel1993spectroscopic}%
  \BibitemOpen
  \bibfield  {author} {\bibinfo {author} {\bibfnamefont {B.}~\bibnamefont
  {Dardel}}, \bibinfo {author} {\bibfnamefont {M.}~\bibnamefont {Grioni}},
  \bibinfo {author} {\bibfnamefont {D.}~\bibnamefont {Malterre}}, \bibinfo
  {author} {\bibfnamefont {P.}~\bibnamefont {Weibel}}, \bibinfo {author}
  {\bibfnamefont {Y.}~\bibnamefont {Baer}},\ and\ \bibinfo {author}
  {\bibfnamefont {F.}~\bibnamefont {Levy}},\ }\bibfield  {title} {\bibinfo
  {title} {{Spectroscopic observation of charge-density-wave-induced changes in
  the electronic structure of 2H-TaSe2}},\ }\href
  {https://doi.org/10.1088/0953-8984/5/33/020} {\bibfield  {journal} {\bibinfo
  {journal} {Journal of Physics: Condensed Matter}\ }\textbf {\bibinfo {volume}
  {5}},\ \bibinfo {pages} {6111} (\bibinfo {year} {1993})}\BibitemShut
  {NoStop}%
\bibitem [{\citenamefont {Straub}\ \emph
  {et~al.}(1999{\natexlab{b}})\citenamefont {Straub}, \citenamefont {Finteis},
  \citenamefont {Claessen}, \citenamefont {Steiner}, \citenamefont {H\"ufner},
  \citenamefont {Blaha}, \citenamefont {Oglesby},\ and\ \citenamefont
  {Bucher}}]{straub1999charge}%
  \BibitemOpen
  \bibfield  {author} {\bibinfo {author} {\bibfnamefont {T.}~\bibnamefont
  {Straub}}, \bibinfo {author} {\bibfnamefont {T.}~\bibnamefont {Finteis}},
  \bibinfo {author} {\bibfnamefont {R.}~\bibnamefont {Claessen}}, \bibinfo
  {author} {\bibfnamefont {P.}~\bibnamefont {Steiner}}, \bibinfo {author}
  {\bibfnamefont {S.}~\bibnamefont {H\"ufner}}, \bibinfo {author}
  {\bibfnamefont {P.}~\bibnamefont {Blaha}}, \bibinfo {author} {\bibfnamefont
  {C.~S.}\ \bibnamefont {Oglesby}},\ and\ \bibinfo {author} {\bibfnamefont
  {E.}~\bibnamefont {Bucher}},\ }\bibfield  {title} {\bibinfo {title}
  {{Charge-Density-Wave Mechanism in
  $2\mathit{H}\ensuremath{-}{\mathrm{NbSe}}_{2}$: Photoemission Results}},\
  }\href {https://doi.org/10.1103/PhysRevLett.82.4504} {\bibfield  {journal}
  {\bibinfo  {journal} {Phys. Rev. Lett.}\ }\textbf {\bibinfo {volume} {82}},\
  \bibinfo {pages} {4504} (\bibinfo {year} {1999}{\natexlab{b}})}\BibitemShut
  {NoStop}%
\bibitem [{\citenamefont {Ruzicka}\ \emph {et~al.}(2001)\citenamefont
  {Ruzicka}, \citenamefont {Degiorgi}, \citenamefont {Berger}, \citenamefont
  {Ga\'al},\ and\ \citenamefont {Forr\'o}}]{ruzicka2001charge}%
  \BibitemOpen
  \bibfield  {author} {\bibinfo {author} {\bibfnamefont {B.}~\bibnamefont
  {Ruzicka}}, \bibinfo {author} {\bibfnamefont {L.}~\bibnamefont {Degiorgi}},
  \bibinfo {author} {\bibfnamefont {H.}~\bibnamefont {Berger}}, \bibinfo
  {author} {\bibfnamefont {R.}~\bibnamefont {Ga\'al}},\ and\ \bibinfo {author}
  {\bibfnamefont {L.}~\bibnamefont {Forr\'o}},\ }\bibfield  {title} {\bibinfo
  {title} {{Charge Dynamics of 2H- ${\mathrm{TaSe}}_{2}$ along the
  Less-Conducting $\mathit{c}$-Axis}},\ }\href
  {https://doi.org/10.1103/PhysRevLett.86.4136} {\bibfield  {journal} {\bibinfo
   {journal} {Phys. Rev. Lett.}\ }\textbf {\bibinfo {volume} {86}},\ \bibinfo
  {pages} {4136} (\bibinfo {year} {2001})}\BibitemShut {NoStop}%
\bibitem [{\citenamefont {Taraphder}\ \emph {et~al.}(2011)\citenamefont
  {Taraphder}, \citenamefont {Koley}, \citenamefont {Vidhyadhiraja},\ and\
  \citenamefont {Laad}}]{taraphder2011preformed}%
  \BibitemOpen
  \bibfield  {author} {\bibinfo {author} {\bibfnamefont {A.}~\bibnamefont
  {Taraphder}}, \bibinfo {author} {\bibfnamefont {S.}~\bibnamefont {Koley}},
  \bibinfo {author} {\bibfnamefont {N.~S.}\ \bibnamefont {Vidhyadhiraja}},\
  and\ \bibinfo {author} {\bibfnamefont {M.~S.}\ \bibnamefont {Laad}},\
  }\bibfield  {title} {\bibinfo {title} {{Preformed Excitonic Liquid Route to a
  Charge Density Wave in
  $2H\mathrm{\text{\ensuremath{-}}}{\mathrm{TaSe}}_{2}$}},\ }\href
  {https://doi.org/10.1103/PhysRevLett.106.236405} {\bibfield  {journal}
  {\bibinfo  {journal} {Phys. Rev. Lett.}\ }\textbf {\bibinfo {volume} {106}},\
  \bibinfo {pages} {236405} (\bibinfo {year} {2011})}\BibitemShut {NoStop}%
\bibitem [{\citenamefont {Koley}\ \emph {et~al.}(2014)\citenamefont {Koley},
  \citenamefont {Laad}, \citenamefont {Vidhyadhiraja},\ and\ \citenamefont
  {Taraphder}}]{koley2014preformed}%
  \BibitemOpen
  \bibfield  {author} {\bibinfo {author} {\bibfnamefont {S.}~\bibnamefont
  {Koley}}, \bibinfo {author} {\bibfnamefont {M.~S.}\ \bibnamefont {Laad}},
  \bibinfo {author} {\bibfnamefont {N.~S.}\ \bibnamefont {Vidhyadhiraja}},\
  and\ \bibinfo {author} {\bibfnamefont {A.}~\bibnamefont {Taraphder}},\
  }\bibfield  {title} {\bibinfo {title} {{Preformed excitons, orbital
  selectivity, and charge density wave order in
  $1T\text{\ensuremath{-}}{\mathrm{TiSe}}_{2}$}},\ }\href
  {https://doi.org/10.1103/PhysRevB.90.115146} {\bibfield  {journal} {\bibinfo
  {journal} {Phys. Rev. B}\ }\textbf {\bibinfo {volume} {90}},\ \bibinfo
  {pages} {115146} (\bibinfo {year} {2014})}\BibitemShut {NoStop}%
\bibitem [{\citenamefont {Zhou}\ \emph {et~al.}(2017)\citenamefont {Zhou},
  \citenamefont {Kanoda},\ and\ \citenamefont {Ng}}]{zhou2017quantum}%
  \BibitemOpen
  \bibfield  {author} {\bibinfo {author} {\bibfnamefont {Y.}~\bibnamefont
  {Zhou}}, \bibinfo {author} {\bibfnamefont {K.}~\bibnamefont {Kanoda}},\ and\
  \bibinfo {author} {\bibfnamefont {T.-K.}\ \bibnamefont {Ng}},\ }\bibfield
  {title} {\bibinfo {title} {Quantum spin liquid states},\ }\href
  {https://doi.org/10.1103/RevModPhys.89.025003} {\bibfield  {journal}
  {\bibinfo  {journal} {Rev. Mod. Phys.}\ }\textbf {\bibinfo {volume} {89}},\
  \bibinfo {pages} {025003} (\bibinfo {year} {2017})}\BibitemShut {NoStop}%
\bibitem [{\citenamefont {Law}\ and\ \citenamefont {Lee}(2017)}]{law20171t}%
  \BibitemOpen
  \bibfield  {author} {\bibinfo {author} {\bibfnamefont {K.~T.}\ \bibnamefont
  {Law}}\ and\ \bibinfo {author} {\bibfnamefont {P.~A.}\ \bibnamefont {Lee}},\
  }\bibfield  {title} {\bibinfo {title} {{1T-TaS2 as a quantum spin liquid}},\
  }\href {https://doi.org/10.1073/pnas.1706769114} {\bibfield  {journal}
  {\bibinfo  {journal} {Proceedings of the National Academy of Sciences}\
  }\textbf {\bibinfo {volume} {114}},\ \bibinfo {pages} {6996} (\bibinfo {year}
  {2017})}\BibitemShut {NoStop}%
\bibitem [{\citenamefont {Anderson}(1973)}]{anderson1973resonating}%
  \BibitemOpen
  \bibfield  {author} {\bibinfo {author} {\bibfnamefont {P.~W.}\ \bibnamefont
  {Anderson}},\ }\bibfield  {title} {\bibinfo {title} {Resonating valence
  bonds: A new kind of insulator?},\ }\href
  {https://doi.org/10.1016/0025-5408(73)90167-0} {\bibfield  {journal}
  {\bibinfo  {journal} {Materials Research Bulletin}\ }\textbf {\bibinfo
  {volume} {8}},\ \bibinfo {pages} {153} (\bibinfo {year} {1973})}\BibitemShut
  {NoStop}%
\bibitem [{\citenamefont {Fazekas}\ and\ \citenamefont
  {Anderson}(1974)}]{fazekas1974ground}%
  \BibitemOpen
  \bibfield  {author} {\bibinfo {author} {\bibfnamefont {P.}~\bibnamefont
  {Fazekas}}\ and\ \bibinfo {author} {\bibfnamefont {P.~W.}\ \bibnamefont
  {Anderson}},\ }\bibfield  {title} {\bibinfo {title} {On the ground state
  properties of the anisotropic triangular antiferromagnet},\ }\href
  {https://doi.org/10.1080/14786439808206568} {\bibfield  {journal} {\bibinfo
  {journal} {Philosophical Magazine}\ }\textbf {\bibinfo {volume} {30}},\
  \bibinfo {pages} {423} (\bibinfo {year} {1974})}\BibitemShut {NoStop}%
\bibitem [{\citenamefont {Zhang}\ \emph {et~al.}(2024)\citenamefont {Zhang},
  \citenamefont {He}, \citenamefont {Zhang}, \citenamefont {Chen},
  \citenamefont {Jia}, \citenamefont {Hou}, \citenamefont {Ji}, \citenamefont
  {Yang}, \citenamefont {Zhang}, \citenamefont {Liu} \emph
  {et~al.}}]{zhang2024quantum}%
  \BibitemOpen
  \bibfield  {author} {\bibinfo {author} {\bibfnamefont {Q.}~\bibnamefont
  {Zhang}}, \bibinfo {author} {\bibfnamefont {W.-Y.}\ \bibnamefont {He}},
  \bibinfo {author} {\bibfnamefont {Y.}~\bibnamefont {Zhang}}, \bibinfo
  {author} {\bibfnamefont {Y.}~\bibnamefont {Chen}}, \bibinfo {author}
  {\bibfnamefont {L.}~\bibnamefont {Jia}}, \bibinfo {author} {\bibfnamefont
  {Y.}~\bibnamefont {Hou}}, \bibinfo {author} {\bibfnamefont {H.}~\bibnamefont
  {Ji}}, \bibinfo {author} {\bibfnamefont {H.}~\bibnamefont {Yang}}, \bibinfo
  {author} {\bibfnamefont {T.}~\bibnamefont {Zhang}}, \bibinfo {author}
  {\bibfnamefont {L.}~\bibnamefont {Liu}}, \emph {et~al.},\ }\bibfield  {title}
  {\bibinfo {title} {{Quantum spin liquid signatures in monolayer
  1T-NbSe$_2$}},\ }\href {https://doi.org/10.1038/s41467-024-46612-1}
  {\bibfield  {journal} {\bibinfo  {journal} {Nature Communications}\ }\textbf
  {\bibinfo {volume} {15}},\ \bibinfo {pages} {2336} (\bibinfo {year}
  {2024})}\BibitemShut {NoStop}%
\bibitem [{\citenamefont {Ruan}\ \emph {et~al.}(2021)\citenamefont {Ruan},
  \citenamefont {Chen}, \citenamefont {Tang}, \citenamefont {Hwang},
  \citenamefont {Tsai}, \citenamefont {Lee}, \citenamefont {Wu}, \citenamefont
  {Ryu}, \citenamefont {Kahn}, \citenamefont {Liou} \emph
  {et~al.}}]{ruan2021evidence}%
  \BibitemOpen
  \bibfield  {author} {\bibinfo {author} {\bibfnamefont {W.}~\bibnamefont
  {Ruan}}, \bibinfo {author} {\bibfnamefont {Y.}~\bibnamefont {Chen}}, \bibinfo
  {author} {\bibfnamefont {S.}~\bibnamefont {Tang}}, \bibinfo {author}
  {\bibfnamefont {J.}~\bibnamefont {Hwang}}, \bibinfo {author} {\bibfnamefont
  {H.-Z.}\ \bibnamefont {Tsai}}, \bibinfo {author} {\bibfnamefont {R.~L.}\
  \bibnamefont {Lee}}, \bibinfo {author} {\bibfnamefont {M.}~\bibnamefont
  {Wu}}, \bibinfo {author} {\bibfnamefont {H.}~\bibnamefont {Ryu}}, \bibinfo
  {author} {\bibfnamefont {S.}~\bibnamefont {Kahn}}, \bibinfo {author}
  {\bibfnamefont {F.}~\bibnamefont {Liou}}, \emph {et~al.},\ }\bibfield
  {title} {\bibinfo {title} {{Evidence for quantum spin liquid behaviour in
  single-layer 1T-TaSe$_2$ from scanning tunnelling microscopy}},\ }\href
  {https://doi.org/10.1038/s41567-021-01321-0} {\bibfield  {journal} {\bibinfo
  {journal} {Nature Physics}\ }\textbf {\bibinfo {volume} {17}},\ \bibinfo
  {pages} {1154} (\bibinfo {year} {2021})}\BibitemShut {NoStop}%
\bibitem [{\citenamefont {Kresse}\ and\ \citenamefont
  {Hafner}(1993)}]{Kresse1}%
  \BibitemOpen
  \bibfield  {author} {\bibinfo {author} {\bibfnamefont {G.}~\bibnamefont
  {Kresse}}\ and\ \bibinfo {author} {\bibfnamefont {J.}~\bibnamefont
  {Hafner}},\ }\bibfield  {title} {\bibinfo {title} {Ab initio molecular
  dynamics for liquid metals},\ }\href
  {https://doi.org/https://doi.org/10.1103/PhysRevB.47.558} {\bibfield
  {journal} {\bibinfo  {journal} {Phys. Rev. B}\ }\textbf {\bibinfo {volume}
  {47}},\ \bibinfo {pages} {558} (\bibinfo {year} {1993})}\BibitemShut
  {NoStop}%
\bibitem [{\citenamefont {Kresse}\ and\ \citenamefont
  {Hafner}(1994)}]{Kresse2}%
  \BibitemOpen
  \bibfield  {author} {\bibinfo {author} {\bibfnamefont {G.}~\bibnamefont
  {Kresse}}\ and\ \bibinfo {author} {\bibfnamefont {J.}~\bibnamefont
  {Hafner}},\ }\bibfield  {title} {\bibinfo {title} {Ab initio
  molecular-dynamics simulation of the liquid-metal–amorphous-semiconductor
  transition in germanium},\ }\href
  {https://doi.org/https://doi.org/10.1103/PhysRevB.49.14251} {\bibfield
  {journal} {\bibinfo  {journal} {Phys. Rev. B}\ }\textbf {\bibinfo {volume}
  {49}},\ \bibinfo {pages} {14251} (\bibinfo {year} {1994})}\BibitemShut
  {NoStop}%
\bibitem [{\citenamefont {Kresse}\ and\ \citenamefont
  {Furthmüller}(1996)}]{Kresse3}%
  \BibitemOpen
  \bibfield  {author} {\bibinfo {author} {\bibfnamefont {G.}~\bibnamefont
  {Kresse}}\ and\ \bibinfo {author} {\bibfnamefont {J.}~\bibnamefont
  {Furthmüller}},\ }\bibfield  {title} {\bibinfo {title} {Efficiency of
  ab-initio total energy calculations for metals and semiconductors using a
  plane-wave basis set},\ }\href@noop {} {\bibfield  {journal} {\bibinfo
  {journal} {Comput. Mat. Sci.}\ }\textbf {\bibinfo {volume} {6}},\ \bibinfo
  {pages} {15} (\bibinfo {year} {1996})}\BibitemShut {NoStop}%
\bibitem [{\citenamefont {Kresse}\ and\ \citenamefont
  {Furthmuller}(1996)}]{Kresse4}%
  \BibitemOpen
  \bibfield  {author} {\bibinfo {author} {\bibfnamefont {G.}~\bibnamefont
  {Kresse}}\ and\ \bibinfo {author} {\bibfnamefont {J.}~\bibnamefont
  {Furthmuller}},\ }\bibfield  {title} {\bibinfo {title} {Efficient iterative
  schemes for ab initio total-energy calculations using a plane-wave basis
  set},\ }\href {https://doi.org/https://doi.org/10.1103/PhysRevB.54.11169}
  {\bibfield  {journal} {\bibinfo  {journal} {Phys. Rev. B}\ }\textbf {\bibinfo
  {volume} {54}},\ \bibinfo {pages} {11169} (\bibinfo {year}
  {1996})}\BibitemShut {NoStop}%
\bibitem [{\citenamefont {Blochl }(1994)}]{Blochl}%
  \BibitemOpen
  \bibfield  {author} {\bibinfo {author} {\bibfnamefont {P.~E.}\ \bibnamefont
  {Blochl }},\ }\bibfield  {title} {\bibinfo {title} {Projector augmented-wave
  method},\ }\href {https://doi.org/https://doi.org/10.1103/PhysRevB.50.17953}
  {\bibfield  {journal} {\bibinfo  {journal} {Phys. Rev. B}\ }\textbf {\bibinfo
  {volume} {50}},\ \bibinfo {pages} {17953} (\bibinfo {year}
  {1994})}\BibitemShut {NoStop}%
\bibitem [{\citenamefont {Kresse}\ and\ \citenamefont
  {Joubert}(1999)}]{Kresse5}%
  \BibitemOpen
  \bibfield  {author} {\bibinfo {author} {\bibfnamefont {G.}~\bibnamefont
  {Kresse}}\ and\ \bibinfo {author} {\bibfnamefont {D.}~\bibnamefont
  {Joubert}},\ }\bibfield  {title} {\bibinfo {title} {From ultrasoft
  pseudopotentials to the projector augmented-wave method},\ }\href
  {https://doi.org/https://doi.org/10.1103/PhysRevB.59.1758} {\bibfield
  {journal} {\bibinfo  {journal} {Phys. Rev. B}\ }\textbf {\bibinfo {volume}
  {59}},\ \bibinfo {pages} {1758} (\bibinfo {year} {1999})}\BibitemShut
  {NoStop}%
\bibitem [{\citenamefont {J.~P.~Perdew}\ and\ \citenamefont
  {Ernzerhof}(1996)}]{Perdew}%
  \BibitemOpen
  \bibfield  {author} {\bibinfo {author} {\bibfnamefont {K.~B.}\ \bibnamefont
  {J.~P.~Perdew}}\ and\ \bibinfo {author} {\bibfnamefont {M.}~\bibnamefont
  {Ernzerhof}},\ }\bibfield  {title} {\bibinfo {title} {Generalized gradient
  approximation made simple},\ }\href
  {https://doi.org/https://doi.org/10.1103/PhysRevLett.77.3865} {\bibfield
  {journal} {\bibinfo  {journal} {Phys. Rev. Lett.}\ }\textbf {\bibinfo
  {volume} {77}},\ \bibinfo {pages} {3865} (\bibinfo {year}
  {1996})}\BibitemShut {NoStop}%
\bibitem [{\citenamefont {J.~P.~Perdew}\ and\ \citenamefont
  {Matthias}(1997)}]{Perdew2}%
  \BibitemOpen
  \bibfield  {author} {\bibinfo {author} {\bibfnamefont {B.~K.}\ \bibnamefont
  {J.~P.~Perdew}}\ and\ \bibinfo {author} {\bibfnamefont {E.}~\bibnamefont
  {Matthias}},\ }\bibfield  {title} {\bibinfo {title} {Generalized gradient
  approximation made simple},\ }\href
  {https://doi.org/https://doi.org/10.1103/PhysRevLett.78.1396} {\bibfield
  {journal} {\bibinfo  {journal} {Phys. Rev. Lett.}\ }\textbf {\bibinfo
  {volume} {78}},\ \bibinfo {pages} {1396} (\bibinfo {year}
  {1997})}\BibitemShut {NoStop}%
\bibitem [{\citenamefont {Grimme}(2006)}]{Grimme}%
  \BibitemOpen
  \bibfield  {author} {\bibinfo {author} {\bibfnamefont {S.}~\bibnamefont
  {Grimme}},\ }\bibfield  {title} {\bibinfo {title} {Semiempirical gga-type
  density functional constructed with a long-range dispersion correction},\
  }\href {https://doi.org/https://doi.org/10.1002/jcc.20495} {\bibfield
  {journal} {\bibinfo  {journal} {Journal of Computational Chemistry}\ }\textbf
  {\bibinfo {volume} {27}},\ \bibinfo {pages} {1787} (\bibinfo {year}
  {2006})}\BibitemShut {NoStop}%
\bibitem [{\citenamefont {Aichhorn}\ \emph {et~al.}(2009)\citenamefont
  {Aichhorn}, \citenamefont {Pourovskii}, \citenamefont {Vildosola},
  \citenamefont {Ferrero}, \citenamefont {Parcollet}, \citenamefont {Miyake},
  \citenamefont {Georges},\ and\ \citenamefont
  {Biermann}}]{aichhorn2009dynamical}%
  \BibitemOpen
  \bibfield  {author} {\bibinfo {author} {\bibfnamefont {M.}~\bibnamefont
  {Aichhorn}}, \bibinfo {author} {\bibfnamefont {L.}~\bibnamefont
  {Pourovskii}}, \bibinfo {author} {\bibfnamefont {V.}~\bibnamefont
  {Vildosola}}, \bibinfo {author} {\bibfnamefont {M.}~\bibnamefont {Ferrero}},
  \bibinfo {author} {\bibfnamefont {O.}~\bibnamefont {Parcollet}}, \bibinfo
  {author} {\bibfnamefont {T.}~\bibnamefont {Miyake}}, \bibinfo {author}
  {\bibfnamefont {A.}~\bibnamefont {Georges}},\ and\ \bibinfo {author}
  {\bibfnamefont {S.}~\bibnamefont {Biermann}},\ }\bibfield  {title} {\bibinfo
  {title} {{Dynamical mean-field theory within an augmented plane-wave
  framework: Assessing electronic correlations in the iron pnictide LaFeAsO}},\
  }\href {https://doi.org/10.1103/PhysRevB.80.085101} {\bibfield  {journal}
  {\bibinfo  {journal} {Phys. Rev. B}\ }\textbf {\bibinfo {volume} {80}},\
  \bibinfo {pages} {085101} (\bibinfo {year} {2009})}\BibitemShut {NoStop}%
\bibitem [{\citenamefont {Aichhorn}\ \emph {et~al.}(2011)\citenamefont
  {Aichhorn}, \citenamefont {Pourovskii},\ and\ \citenamefont
  {Georges}}]{aichhorn2011importance}%
  \BibitemOpen
  \bibfield  {author} {\bibinfo {author} {\bibfnamefont {M.}~\bibnamefont
  {Aichhorn}}, \bibinfo {author} {\bibfnamefont {L.}~\bibnamefont
  {Pourovskii}},\ and\ \bibinfo {author} {\bibfnamefont {A.}~\bibnamefont
  {Georges}},\ }\bibfield  {title} {\bibinfo {title} {Importance of electronic
  correlations for structural and magnetic properties of the iron pnictide
  superconductor lafeaso},\ }\href {https://doi.org/10.1103/PhysRevB.84.054529}
  {\bibfield  {journal} {\bibinfo  {journal} {Phys. Rev. B}\ }\textbf {\bibinfo
  {volume} {84}},\ \bibinfo {pages} {054529} (\bibinfo {year}
  {2011})}\BibitemShut {NoStop}%
\bibitem [{\citenamefont {Aichhorn}\ \emph {et~al.}(2016)\citenamefont
  {Aichhorn}, \citenamefont {Pourovskii}, \citenamefont {Seth}, \citenamefont
  {Vildosola}, \citenamefont {Zingl}, \citenamefont {Peil}, \citenamefont
  {Deng}, \citenamefont {Mravlje}, \citenamefont {Kraberger}, \citenamefont
  {Martins} \emph {et~al.}}]{aichhorn2016triqs}%
  \BibitemOpen
  \bibfield  {author} {\bibinfo {author} {\bibfnamefont {M.}~\bibnamefont
  {Aichhorn}}, \bibinfo {author} {\bibfnamefont {L.}~\bibnamefont
  {Pourovskii}}, \bibinfo {author} {\bibfnamefont {P.}~\bibnamefont {Seth}},
  \bibinfo {author} {\bibfnamefont {V.}~\bibnamefont {Vildosola}}, \bibinfo
  {author} {\bibfnamefont {M.}~\bibnamefont {Zingl}}, \bibinfo {author}
  {\bibfnamefont {O.~E.}\ \bibnamefont {Peil}}, \bibinfo {author}
  {\bibfnamefont {X.}~\bibnamefont {Deng}}, \bibinfo {author} {\bibfnamefont
  {J.}~\bibnamefont {Mravlje}}, \bibinfo {author} {\bibfnamefont {G.~J.}\
  \bibnamefont {Kraberger}}, \bibinfo {author} {\bibfnamefont {C.}~\bibnamefont
  {Martins}}, \emph {et~al.},\ }\bibfield  {title} {\bibinfo {title}
  {{TRIQS/DFTTools: A TRIQS application for ab initio calculations of
  correlated materials}},\ }\href {https://doi.org/10.1016/j.cpc.2016.03.014}
  {\bibfield  {journal} {\bibinfo  {journal} {Computer Physics Communications}\
  }\textbf {\bibinfo {volume} {204}},\ \bibinfo {pages} {200} (\bibinfo {year}
  {2016})}\BibitemShut {NoStop}%
\bibitem [{mer(2022)}]{merkel2022soliddmft}%
  \BibitemOpen
  \bibfield  {title} {\bibinfo {title} {{solid\_dmft: gray-boxing DFT+DMFT
  materials simulations with TRIQS}},\ }\href
  {https://doi.org/10.21105/joss.04623} {\bibfield  {journal} {\bibinfo
  {journal} {Journal of Open Source Software}\ }\textbf {\bibinfo {volume}
  {7}},\ \bibinfo {pages} {4623} (\bibinfo {year} {2022})}\BibitemShut
  {NoStop}%
\bibitem [{\citenamefont {Parcollet}\ \emph {et~al.}(2015)\citenamefont
  {Parcollet}, \citenamefont {Ferrero}, \citenamefont {Ayral}, \citenamefont
  {Hafermann}, \citenamefont {Krivenko}, \citenamefont {Messio},\ and\
  \citenamefont {Seth}}]{parcollet2015triqs}%
  \BibitemOpen
  \bibfield  {author} {\bibinfo {author} {\bibfnamefont {O.}~\bibnamefont
  {Parcollet}}, \bibinfo {author} {\bibfnamefont {M.}~\bibnamefont {Ferrero}},
  \bibinfo {author} {\bibfnamefont {T.}~\bibnamefont {Ayral}}, \bibinfo
  {author} {\bibfnamefont {H.}~\bibnamefont {Hafermann}}, \bibinfo {author}
  {\bibfnamefont {I.}~\bibnamefont {Krivenko}}, \bibinfo {author}
  {\bibfnamefont {L.}~\bibnamefont {Messio}},\ and\ \bibinfo {author}
  {\bibfnamefont {P.}~\bibnamefont {Seth}},\ }\bibfield  {title} {\bibinfo
  {title} {{TRIQS: A toolbox for research on interacting quantum systems}},\
  }\href {https://doi.org/10.1016/j.cpc.2015.04.023} {\bibfield  {journal}
  {\bibinfo  {journal} {Computer Physics Communications}\ }\textbf {\bibinfo
  {volume} {196}},\ \bibinfo {pages} {398} (\bibinfo {year}
  {2015})}\BibitemShut {NoStop}%
\bibitem [{\citenamefont {Werner}\ and\ \citenamefont
  {Millis}(2006)}]{werner2006hybridization}%
  \BibitemOpen
  \bibfield  {author} {\bibinfo {author} {\bibfnamefont {P.}~\bibnamefont
  {Werner}}\ and\ \bibinfo {author} {\bibfnamefont {A.~J.}\ \bibnamefont
  {Millis}},\ }\bibfield  {title} {\bibinfo {title} {Hybridization expansion
  impurity solver: General formulation and application to kondo lattice and
  two-orbital models},\ }\href {https://doi.org/10.1103/PhysRevB.74.155107}
  {\bibfield  {journal} {\bibinfo  {journal} {Phys. Rev. B}\ }\textbf {\bibinfo
  {volume} {74}},\ \bibinfo {pages} {155107} (\bibinfo {year}
  {2006})}\BibitemShut {NoStop}%
\bibitem [{\citenamefont {Seth}\ \emph {et~al.}(2016)\citenamefont {Seth},
  \citenamefont {Krivenko}, \citenamefont {Ferrero},\ and\ \citenamefont
  {Parcollet}}]{seth2016triqs}%
  \BibitemOpen
  \bibfield  {author} {\bibinfo {author} {\bibfnamefont {P.}~\bibnamefont
  {Seth}}, \bibinfo {author} {\bibfnamefont {I.}~\bibnamefont {Krivenko}},
  \bibinfo {author} {\bibfnamefont {M.}~\bibnamefont {Ferrero}},\ and\ \bibinfo
  {author} {\bibfnamefont {O.}~\bibnamefont {Parcollet}},\ }\bibfield  {title}
  {\bibinfo {title} {Triqs/cthyb: A continuous-time quantum monte carlo
  hybridisation expansion solver for quantum impurity problems},\ }\href
  {https://doi.org/10.1016/j.cpc.2015.10.023} {\bibfield  {journal} {\bibinfo
  {journal} {Computer Physics Communications}\ }\textbf {\bibinfo {volume}
  {200}},\ \bibinfo {pages} {274} (\bibinfo {year} {2016})}\BibitemShut
  {NoStop}%
\bibitem [{\citenamefont {Vaugier}\ \emph {et~al.}(2012)\citenamefont
  {Vaugier}, \citenamefont {Jiang},\ and\ \citenamefont
  {Biermann}}]{vaugier2012hubbard}%
  \BibitemOpen
  \bibfield  {author} {\bibinfo {author} {\bibfnamefont {L.}~\bibnamefont
  {Vaugier}}, \bibinfo {author} {\bibfnamefont {H.}~\bibnamefont {Jiang}},\
  and\ \bibinfo {author} {\bibfnamefont {S.}~\bibnamefont {Biermann}},\
  }\bibfield  {title} {\bibinfo {title} {Hubbard $u$ and hund exchange $j$ in
  transition metal oxides: Screening versus localization trends from
  constrained random phase approximation},\ }\href
  {https://doi.org/10.1103/PhysRevB.86.165105} {\bibfield  {journal} {\bibinfo
  {journal} {Phys. Rev. B}\ }\textbf {\bibinfo {volume} {86}},\ \bibinfo
  {pages} {165105} (\bibinfo {year} {2012})}\BibitemShut {NoStop}%
\bibitem [{\citenamefont {Kraberger}\ \emph {et~al.}(2017)\citenamefont
  {Kraberger}, \citenamefont {Triebl}, \citenamefont {Zingl},\ and\
  \citenamefont {Aichhorn}}]{kraberger2017maximum}%
  \BibitemOpen
  \bibfield  {author} {\bibinfo {author} {\bibfnamefont {G.~J.}\ \bibnamefont
  {Kraberger}}, \bibinfo {author} {\bibfnamefont {R.}~\bibnamefont {Triebl}},
  \bibinfo {author} {\bibfnamefont {M.}~\bibnamefont {Zingl}},\ and\ \bibinfo
  {author} {\bibfnamefont {M.}~\bibnamefont {Aichhorn}},\ }\bibfield  {title}
  {\bibinfo {title} {{Maximum entropy formalism for the analytic continuation
  of matrix-valued Green's functions}},\ }\href
  {https://doi.org/10.1103/PhysRevB.96.155128} {\bibfield  {journal} {\bibinfo
  {journal} {Phys. Rev. B}\ }\textbf {\bibinfo {volume} {96}},\ \bibinfo
  {pages} {155128} (\bibinfo {year} {2017})}\BibitemShut {NoStop}%
\bibitem [{\citenamefont {Liu}\ \emph {et~al.}(2021)\citenamefont {Liu},
  \citenamefont {Yang}, \citenamefont {Huang}, \citenamefont {Song},
  \citenamefont {Zhang}, \citenamefont {Huang}, \citenamefont {Hou},
  \citenamefont {Chen}, \citenamefont {Xu}, \citenamefont {Zhang} \emph
  {et~al.}}]{liu2021direct}%
  \BibitemOpen
  \bibfield  {author} {\bibinfo {author} {\bibfnamefont {L.}~\bibnamefont
  {Liu}}, \bibinfo {author} {\bibfnamefont {H.}~\bibnamefont {Yang}}, \bibinfo
  {author} {\bibfnamefont {Y.}~\bibnamefont {Huang}}, \bibinfo {author}
  {\bibfnamefont {X.}~\bibnamefont {Song}}, \bibinfo {author} {\bibfnamefont
  {Q.}~\bibnamefont {Zhang}}, \bibinfo {author} {\bibfnamefont
  {Z.}~\bibnamefont {Huang}}, \bibinfo {author} {\bibfnamefont
  {Y.}~\bibnamefont {Hou}}, \bibinfo {author} {\bibfnamefont {Y.}~\bibnamefont
  {Chen}}, \bibinfo {author} {\bibfnamefont {Z.}~\bibnamefont {Xu}}, \bibinfo
  {author} {\bibfnamefont {T.}~\bibnamefont {Zhang}}, \emph {et~al.},\
  }\bibfield  {title} {\bibinfo {title} {{Direct identification of Mott Hubbard
  band pattern beyond charge density wave superlattice in monolayer
  1T-NbSe$_2$}},\ }\href {https://doi.org/10.1038/s41467-021-22233-w}
  {\bibfield  {journal} {\bibinfo  {journal} {Nature communications}\ }\textbf
  {\bibinfo {volume} {12}},\ \bibinfo {pages} {1978} (\bibinfo {year}
  {2021})}\BibitemShut {NoStop}%
\bibitem [{\citenamefont {Chen}\ \emph {et~al.}(2020)\citenamefont {Chen},
  \citenamefont {Ruan}, \citenamefont {Wu}, \citenamefont {Tang}, \citenamefont
  {Ryu}, \citenamefont {Tsai}, \citenamefont {Lee}, \citenamefont {Kahn},
  \citenamefont {Liou}, \citenamefont {Jia} \emph {et~al.}}]{chen2020strong}%
  \BibitemOpen
  \bibfield  {author} {\bibinfo {author} {\bibfnamefont {Y.}~\bibnamefont
  {Chen}}, \bibinfo {author} {\bibfnamefont {W.}~\bibnamefont {Ruan}}, \bibinfo
  {author} {\bibfnamefont {M.}~\bibnamefont {Wu}}, \bibinfo {author}
  {\bibfnamefont {S.}~\bibnamefont {Tang}}, \bibinfo {author} {\bibfnamefont
  {H.}~\bibnamefont {Ryu}}, \bibinfo {author} {\bibfnamefont {H.-Z.}\
  \bibnamefont {Tsai}}, \bibinfo {author} {\bibfnamefont {R.~L.}\ \bibnamefont
  {Lee}}, \bibinfo {author} {\bibfnamefont {S.}~\bibnamefont {Kahn}}, \bibinfo
  {author} {\bibfnamefont {F.}~\bibnamefont {Liou}}, \bibinfo {author}
  {\bibfnamefont {C.}~\bibnamefont {Jia}}, \emph {et~al.},\ }\bibfield  {title}
  {\bibinfo {title} {Strong correlations and orbital texture in single-layer
  1t-tase2},\ }\href {https://doi.org/10.1038/s41567-019-0744-9} {\bibfield
  {journal} {\bibinfo  {journal} {Nature Physics}\ }\textbf {\bibinfo {volume}
  {16}},\ \bibinfo {pages} {218} (\bibinfo {year} {2020})}\BibitemShut
  {NoStop}%
\bibitem [{\citenamefont {Marzari}\ and\ \citenamefont
  {Vanderbilt}(1997)}]{marzari1997maximally}%
  \BibitemOpen
  \bibfield  {author} {\bibinfo {author} {\bibfnamefont {N.}~\bibnamefont
  {Marzari}}\ and\ \bibinfo {author} {\bibfnamefont {D.}~\bibnamefont
  {Vanderbilt}},\ }\bibfield  {title} {\bibinfo {title} {{Maximally localized
  generalized Wannier functions for composite energy bands}},\ }\href
  {https://doi.org/10.1103/PhysRevB.56.12847} {\bibfield  {journal} {\bibinfo
  {journal} {Phys. Rev. B}\ }\textbf {\bibinfo {volume} {56}},\ \bibinfo
  {pages} {12847} (\bibinfo {year} {1997})}\BibitemShut {NoStop}%
\bibitem [{\citenamefont {Pizzi}\ \emph {et~al.}(2020)\citenamefont {Pizzi},
  \citenamefont {Vitale}, \citenamefont {Arita}, \citenamefont {Bl{\"u}gel},
  \citenamefont {Freimuth}, \citenamefont {G{\'e}ranton}, \citenamefont
  {Gibertini}, \citenamefont {Gresch}, \citenamefont {Johnson}, \citenamefont
  {Koretsune} \emph {et~al.}}]{pizzi2020wannier90}%
  \BibitemOpen
  \bibfield  {author} {\bibinfo {author} {\bibfnamefont {G.}~\bibnamefont
  {Pizzi}}, \bibinfo {author} {\bibfnamefont {V.}~\bibnamefont {Vitale}},
  \bibinfo {author} {\bibfnamefont {R.}~\bibnamefont {Arita}}, \bibinfo
  {author} {\bibfnamefont {S.}~\bibnamefont {Bl{\"u}gel}}, \bibinfo {author}
  {\bibfnamefont {F.}~\bibnamefont {Freimuth}}, \bibinfo {author}
  {\bibfnamefont {G.}~\bibnamefont {G{\'e}ranton}}, \bibinfo {author}
  {\bibfnamefont {M.}~\bibnamefont {Gibertini}}, \bibinfo {author}
  {\bibfnamefont {D.}~\bibnamefont {Gresch}}, \bibinfo {author} {\bibfnamefont
  {C.}~\bibnamefont {Johnson}}, \bibinfo {author} {\bibfnamefont
  {T.}~\bibnamefont {Koretsune}}, \emph {et~al.},\ }\bibfield  {title}
  {\bibinfo {title} {Wannier90 as a community code: new features and
  applications},\ }\href {https://doi.org/10.1088/1361-648X/ab51ff} {\bibfield
  {journal} {\bibinfo  {journal} {Journal of Physics: Condensed Matter}\
  }\textbf {\bibinfo {volume} {32}},\ \bibinfo {pages} {165902} (\bibinfo
  {year} {2020})}\BibitemShut {NoStop}%
\bibitem [{\citenamefont {Giannozzi}\ \emph {et~al.}(2009)\citenamefont
  {Giannozzi}, \citenamefont {Baroni}, \citenamefont {Bonini}, \citenamefont
  {Calandra}, \citenamefont {Car}, \citenamefont {Cavazzoni}, \citenamefont
  {Ceresoli}, \citenamefont {Chiarotti}, \citenamefont {Cococcioni},
  \citenamefont {Dabo} \emph {et~al.}}]{giannozzi2009quantum}%
  \BibitemOpen
  \bibfield  {author} {\bibinfo {author} {\bibfnamefont {P.}~\bibnamefont
  {Giannozzi}}, \bibinfo {author} {\bibfnamefont {S.}~\bibnamefont {Baroni}},
  \bibinfo {author} {\bibfnamefont {N.}~\bibnamefont {Bonini}}, \bibinfo
  {author} {\bibfnamefont {M.}~\bibnamefont {Calandra}}, \bibinfo {author}
  {\bibfnamefont {R.}~\bibnamefont {Car}}, \bibinfo {author} {\bibfnamefont
  {C.}~\bibnamefont {Cavazzoni}}, \bibinfo {author} {\bibfnamefont
  {D.}~\bibnamefont {Ceresoli}}, \bibinfo {author} {\bibfnamefont {G.~L.}\
  \bibnamefont {Chiarotti}}, \bibinfo {author} {\bibfnamefont {M.}~\bibnamefont
  {Cococcioni}}, \bibinfo {author} {\bibfnamefont {I.}~\bibnamefont {Dabo}},
  \emph {et~al.},\ }\bibfield  {title} {\bibinfo {title} {{QUANTUM ESPRESSO: a
  modular and open-source software project for quantum simulations of
  materials}},\ }\href {https://doi.org/10.1088/0953-8984/21/39/395502}
  {\bibfield  {journal} {\bibinfo  {journal} {Journal of physics: Condensed
  matter}\ }\textbf {\bibinfo {volume} {21}},\ \bibinfo {pages} {395502}
  (\bibinfo {year} {2009})}\BibitemShut {NoStop}%
\bibitem [{\citenamefont {Giannozzi}\ \emph {et~al.}(2017)\citenamefont
  {Giannozzi}, \citenamefont {Andreussi}, \citenamefont {Brumme}, \citenamefont
  {Bunau}, \citenamefont {Nardelli}, \citenamefont {Calandra}, \citenamefont
  {Car}, \citenamefont {Cavazzoni}, \citenamefont {Ceresoli}, \citenamefont
  {Cococcioni} \emph {et~al.}}]{giannozzi2017advanced}%
  \BibitemOpen
  \bibfield  {author} {\bibinfo {author} {\bibfnamefont {P.}~\bibnamefont
  {Giannozzi}}, \bibinfo {author} {\bibfnamefont {O.}~\bibnamefont
  {Andreussi}}, \bibinfo {author} {\bibfnamefont {T.}~\bibnamefont {Brumme}},
  \bibinfo {author} {\bibfnamefont {O.}~\bibnamefont {Bunau}}, \bibinfo
  {author} {\bibfnamefont {M.~B.}\ \bibnamefont {Nardelli}}, \bibinfo {author}
  {\bibfnamefont {M.}~\bibnamefont {Calandra}}, \bibinfo {author}
  {\bibfnamefont {R.}~\bibnamefont {Car}}, \bibinfo {author} {\bibfnamefont
  {C.}~\bibnamefont {Cavazzoni}}, \bibinfo {author} {\bibfnamefont
  {D.}~\bibnamefont {Ceresoli}}, \bibinfo {author} {\bibfnamefont
  {M.}~\bibnamefont {Cococcioni}}, \emph {et~al.},\ }\bibfield  {title}
  {\bibinfo {title} {Advanced capabilities for materials modelling with quantum
  espresso},\ }\href {https://doi.org/10.1088/1361-648X/aa8f79} {\bibfield
  {journal} {\bibinfo  {journal} {Journal of physics: Condensed matter}\
  }\textbf {\bibinfo {volume} {29}},\ \bibinfo {pages} {465901} (\bibinfo
  {year} {2017})}\BibitemShut {NoStop}%
\bibitem [{\citenamefont {Giannozzi}\ \emph {et~al.}(2020)\citenamefont
  {Giannozzi}, \citenamefont {Baseggio}, \citenamefont {Bonf{\`a}},
  \citenamefont {Brunato}, \citenamefont {Car}, \citenamefont {Carnimeo},
  \citenamefont {Cavazzoni}, \citenamefont {De~Gironcoli}, \citenamefont
  {Delugas}, \citenamefont {Ferrari~Ruffino} \emph
  {et~al.}}]{giannozzi2020quantum}%
  \BibitemOpen
  \bibfield  {author} {\bibinfo {author} {\bibfnamefont {P.}~\bibnamefont
  {Giannozzi}}, \bibinfo {author} {\bibfnamefont {O.}~\bibnamefont {Baseggio}},
  \bibinfo {author} {\bibfnamefont {P.}~\bibnamefont {Bonf{\`a}}}, \bibinfo
  {author} {\bibfnamefont {D.}~\bibnamefont {Brunato}}, \bibinfo {author}
  {\bibfnamefont {R.}~\bibnamefont {Car}}, \bibinfo {author} {\bibfnamefont
  {I.}~\bibnamefont {Carnimeo}}, \bibinfo {author} {\bibfnamefont
  {C.}~\bibnamefont {Cavazzoni}}, \bibinfo {author} {\bibfnamefont
  {S.}~\bibnamefont {De~Gironcoli}}, \bibinfo {author} {\bibfnamefont
  {P.}~\bibnamefont {Delugas}}, \bibinfo {author} {\bibfnamefont
  {F.}~\bibnamefont {Ferrari~Ruffino}}, \emph {et~al.},\ }\bibfield  {title}
  {\bibinfo {title} {Quantum espresso toward the exascale},\ }\href
  {https://doi.org/10.1063/5.0005082} {\bibfield  {journal} {\bibinfo
  {journal} {The Journal of chemical physics}\ }\textbf {\bibinfo {volume}
  {152}},\ \bibinfo {pages} {154105} (\bibinfo {year} {2020})}\BibitemShut
  {NoStop}%
\bibitem [{\citenamefont {Anisimov}\ \emph {et~al.}(2002)\citenamefont
  {Anisimov}, \citenamefont {Nekrasov}, \citenamefont {Kondakov}, \citenamefont
  {Rice},\ and\ \citenamefont {Sigrist}}]{anisimov2002orbital}%
  \BibitemOpen
  \bibfield  {author} {\bibinfo {author} {\bibfnamefont {V.}~\bibnamefont
  {Anisimov}}, \bibinfo {author} {\bibfnamefont {I.}~\bibnamefont {Nekrasov}},
  \bibinfo {author} {\bibfnamefont {D.}~\bibnamefont {Kondakov}}, \bibinfo
  {author} {\bibfnamefont {T.}~\bibnamefont {Rice}},\ and\ \bibinfo {author}
  {\bibfnamefont {M.}~\bibnamefont {Sigrist}},\ }\bibfield  {title} {\bibinfo
  {title} {{Orbital-selective Mott-insulator transition in Ca$_{2-x}$ Sr$_x$
  RuO$_4$}},\ }\href {https://doi.org/10.1140/epjb/e20020021} {\bibfield
  {journal} {\bibinfo  {journal} {The European Physical Journal B-Condensed
  Matter and Complex Systems}\ }\textbf {\bibinfo {volume} {25}},\ \bibinfo
  {pages} {191} (\bibinfo {year} {2002})}\BibitemShut {NoStop}%
\bibitem [{\citenamefont {Vojta}(2010)}]{vojta2010orbital}%
  \BibitemOpen
  \bibfield  {author} {\bibinfo {author} {\bibfnamefont {M.}~\bibnamefont
  {Vojta}},\ }\bibfield  {title} {\bibinfo {title} {Orbital-selective mott
  transitions: Heavy fermions and beyond},\ }\href
  {https://doi.org/10.1007/s10909-010-0206-3} {\bibfield  {journal} {\bibinfo
  {journal} {Journal of Low Temperature Physics}\ }\textbf {\bibinfo {volume}
  {161}},\ \bibinfo {pages} {203} (\bibinfo {year} {2010})}\BibitemShut
  {NoStop}%
\bibitem [{\citenamefont {Boehnke}\ \emph {et~al.}(2016)\citenamefont
  {Boehnke}, \citenamefont {Nilsson}, \citenamefont {Aryasetiawan},\ and\
  \citenamefont {Werner}}]{boehnke2016when}%
  \BibitemOpen
  \bibfield  {author} {\bibinfo {author} {\bibfnamefont {L.}~\bibnamefont
  {Boehnke}}, \bibinfo {author} {\bibfnamefont {F.}~\bibnamefont {Nilsson}},
  \bibinfo {author} {\bibfnamefont {F.}~\bibnamefont {Aryasetiawan}},\ and\
  \bibinfo {author} {\bibfnamefont {P.}~\bibnamefont {Werner}},\ }\bibfield
  {title} {\bibinfo {title} {{When strong correlations become weak: Consistent
  merging of $GW$ and DMFT}},\ }\href
  {https://doi.org/10.1103/PhysRevB.94.201106} {\bibfield  {journal} {\bibinfo
  {journal} {Phys. Rev. B}\ }\textbf {\bibinfo {volume} {94}},\ \bibinfo
  {pages} {201106} (\bibinfo {year} {2016})}\BibitemShut {NoStop}%
\end{thebibliography}%

\end{document}